\documentclass[journal=jctc,manuscript=article]{achemso}
\usepackage{graphicx}
\usepackage{xcolor}
\usepackage{amsmath, amsfonts, amssymb}
\usepackage{bm}
\usepackage{gensymb }
\usepackage{xr}
\usepackage{chemformula}
\usepackage{multirow}
\makeatletter

\newcommand{\red}[1]{#1}

\newcommand{\RQM}{\mathbf{R}^\mathrm{QM}}
\newcommand{\RMM}{\mathbf{R}^\mathrm{MM}}
\newcommand{\density}{\rho(\mathbf{r})}
\newcommand{\dd}{\mathrm{d}}
\newcommand{\rr}{\mathbf{r}}
\newcommand{\I}{\mathrm{I}}
\newcommand{\II}{\mathrm{II}}

\newcommand*{\addFileDependency}[1]{
\typeout{(#1)}
\@addtofilelist{#1}
\IfFileExists{#1}{}{\typeout{No file #1.}}
}\makeatother

\newcommand*{\myexternaldocument}[1]{%
\externaldocument{#1}%
\addFileDependency{#1.tex}%
\addFileDependency{#1.aux}%
}

\myexternaldocument{supplement}

\title{Accurate QM/MM Molecular Dynamics for Periodic Systems in \textsc{GPU4PySCF} with Applications to Enzyme Catalysis}

\author{Chenghan Li}
\author{Garnet Kin-Lic Chan}
\affiliation[Caltech]
{Division of Chemistry and Chemical Engineering, 
California Institute of Technology, Pasadena, California 91125, USA}
\email{gkc1000@gmail.com}

\keywords{QM/MM, molecular dynamics, quantum chemistry}

\begin{document}

\begin{tocentry}
\includegraphics[width=3.25in]{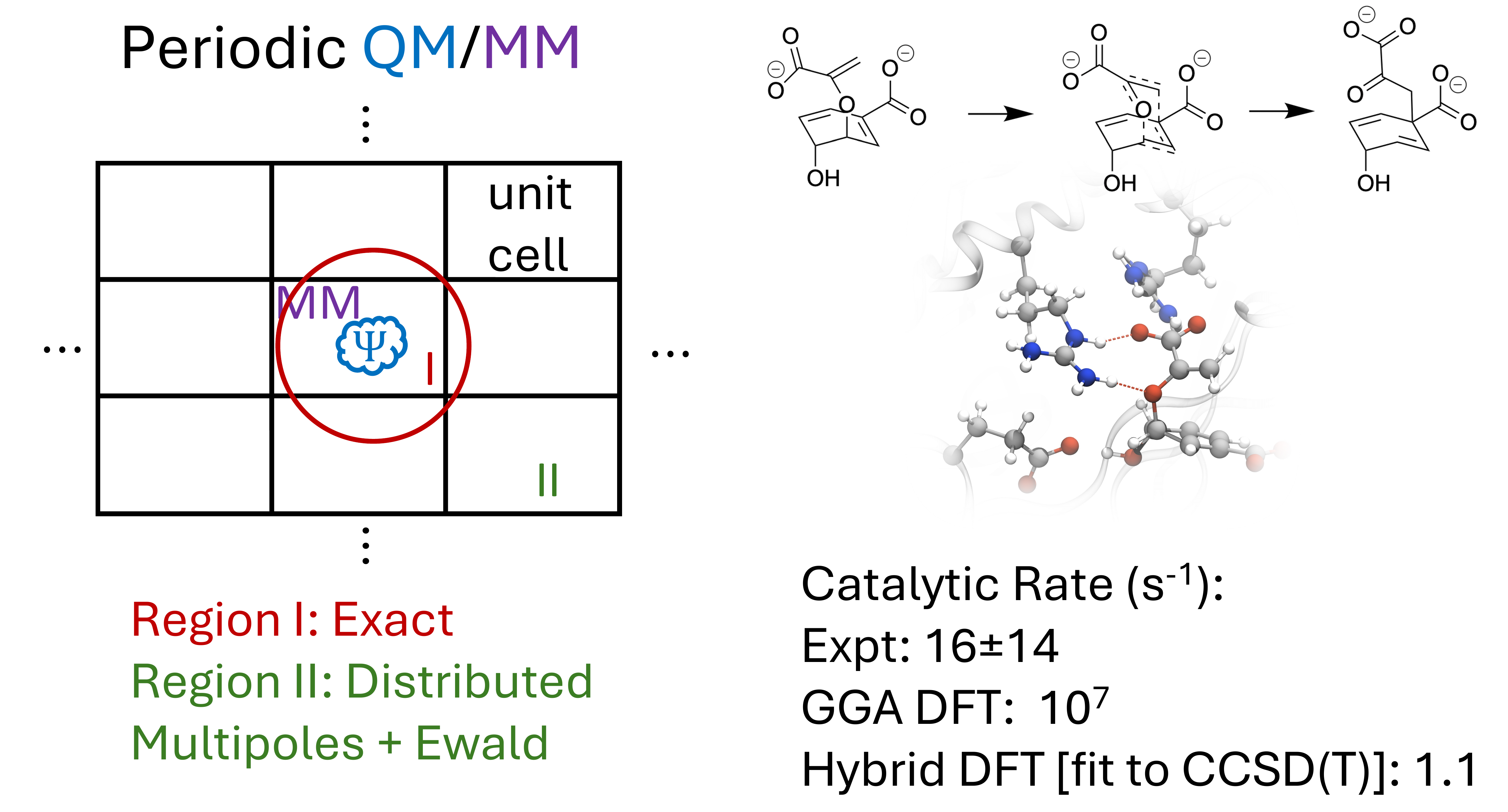}
\end{tocentry}

\begin{abstract}

We present an implementation of the quantum mechanics/molecular mechanics (QM/MM) method for periodic systems using GPU accelerated QM methods, a distributed multipole formulation of the electrostatics, and a pseudo-bond treatment of the QM/MM boundary.  
We demonstrate that our method has well-controlled errors, stable self-consistent QM convergence, and energy-conserving dynamics. 
We further describe an application to the catalytic kinetics of chorismate mutase. Using an accurate hybrid functional reparametrized to coupled cluster energetics, our QM/MM simulations
highlight the sensitivity in the calculated rate to the choice of quantum method, quantum region selection, and local protein conformation.  Our work is provided through the open-source \textsc{PySCF} package using acceleration from the \textsc{GPU4PySCF} module.
\end{abstract}

\maketitle
\section{Introduction}
Quantum mechanics (QM) is the underlying physical theory of chemical phenomena. However, applying a full QM treatment to large-scale chemical problems is both unnecessary and computationally prohibitive. Pioneered by Warshel, Levitt, and Karplus\cite{warshel1976theoretical,karplus1990combined}, the hybrid Quantum Mechanics/Molecular Mechanics (QM/MM) approach addresses this by limiting the QM treatment to critical regions of interest while modeling the surroundings with more approximate empirical force fields\cite{gao1996methods,hu2008free,chung2015oniom,senn2009qm,brunk2015mixed,lonsdale2014qm,cui2021biomolecular,kubavr2023hybrid}.


In this work, we describe and apply a new QM/MM implementation.
Given a QM/MM system partitioning, the factors that define the QM/MM treatment are (i) the choice of method to use in the QM calculation, (ii) the treatment of the electrostatic interaction between the QM and MM regions, (iii) the handling of the QM/MM boundary, and (iv) the quality of the MM force-field.
We present a GPU-based QM/MM implementation with the following characteristics (i) fast algorithms for accurate QM methods, such as hybrid density functionals, in the QM region, (ii)  a multipole representation and self-consistent treatment of the periodic quantum electrostatic potential (denoted here the QM/MM-Multipole method), (iii) a pseudo-bond treatment of the QM/MM boundary, and an open-source implementation within \textsc{PySCF} taking advantage of the GPU accelerated \textsc{GPU4PySCF} module\cite{li2024introducting,wu2024enhancing}, thus (iv) enabling straightforward integration with standard open-source force-fields. 

We carefully benchmark the convergence of the QM/MM-Multipole method (and its resulting effect on the QM self-consistent field stability and energy conservation) in a QM/MM water model and a model for microtubule-mediated GTP hydrolysis. Finally, we investigate the utility of our implementation in enzyme reactions in the context of the catalytic kinetics of chorismate mutase, which has recently been studied at the QM/MM level using a pure density functional description of the QM region~\cite{rak2015brush}.
In our simulations, we are able to employ a more accurate QM method obtained by reparametrizing a hybrid density functional to local coupled cluster energetics. Our resulting reaction rate, extracted by enhanced sampling of the QM/MM potential energy surface, improves the agreement with the experimental estimate, and overall our study highlights the sensitivity of the rate to many aspects of the QM/MM treatment.

\section{Theory and Methods}

\subsection{Energy partitioning}

We consider a total system with periodic boundary conditions consisting of two subsystems, partitioned as two sets of atoms. One subsystem is then described by quantum chemistry methods (in this work, some form of density functional theory) while the other subsystem is described by empirical classical force fields. For the definition of the total energy, we use the electrostatic embedding scheme, where the QM subsystem interacts with non-polarizable MM atoms via a classical Coulomb interaction. Specifically, the  system energy functional is the sum of three components,
\begin{equation} \label{eq:qmmm_energy}
    E(\Psi, \RQM, \RMM) = E_\mathrm{QM}(\Psi, \RQM) + E_\mathrm{QM-MM}(\density, \RMM)
    + E_\mathrm{MM}(\RQM, \RMM)
\end{equation}
where $E_\mathrm{QM}$ is the energy functional of the (periodic) QM subsystem without the MM field (i.e. it is a function of the electronic wavefunction $\Psi$ and the QM atom positions $\RQM$), $E_\mathrm{MM}$ is the energy of the (periodically repeated) MM atoms themselves plus the van der Waals interactions between the QM and MM atoms, both of which are described at the MM level, and $E_\mathrm{QM-MM}$ denotes the electrostatic coupling between the QM electron density $\rho(\mathbf{r})$ and the charges centered on the MM atoms. We minimize this functional with respect to $\Psi$, which relaxes the QM density in the presence of the MM charges and produces a self-consistent treatment of the electrostatic QM/MM coupling. We then obtain the energy and gradients in order to carry out ab initio QM/MM molecular dynamics (MD).

\begin{figure}
    \centering
    \includegraphics{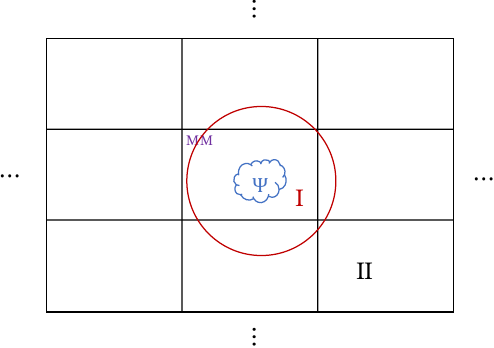}
    \caption{An illustration of the partitioning of the QM-MM electrostatics into near-field (region I) and far-field (region II) components. Each rectangle represents one unit cell. Region I is shown as a sphere centered on the QM subsystem and region II contains the remaining infinite lattice.}
    \label{fig:partitioning}
\end{figure}

\subsection{Electrostatic treatment: background}

In condensed phase systems, a critical component of the functional in Eq.~\ref{eq:qmmm_energy} is the electrostatic coupling, in particular the contributions that arise from long-range periodic electrostatics within each of the three terms. 
It has long been recognized that serious artifacts arise in molecular simulations from ignoring the long-range tail of the electrostatics\cite{brooks1985structural,york1995toward}.

Long-range electrostatics $E_\mathrm{MM}$ are part of standard classical MD methodology and are handled in the force-field part of the QM/MM implementation thus we do not discuss them further here. The electrostatic contributions from the QM charge density in the first two terms of Eq.~\ref{eq:qmmm_energy}, however, require a different treatment, due to the continuous nature of the QM charge density.
Depending on the formulation, the computation of electrostatics involving this charge density can be quite expensive. 
To motivate our formulation, we briefly summarize some background on how other QM/MM implementations treat the periodic electrostatic coupling between the QM density and MM charges, before discussing our formulation in the next section. (Because we will treat the QM/QM electrostatic interaction in the same way as the QM/MM electrostatic interaction, we will not separately survey the QM/QM electrostatic treatments in the literature).


One category of methods represents the electrostatic potential of the periodic MM charge distribution 
on a grid constructed within the reference cell. 
The MM electrostatic potential can then be integrated with the QM charge density in the reference cell\cite{sanz2011efficient,kawashima2019ab}. 
Various kinds of grids and approximations to accelerate this computation have been explored~\cite{sanz2011efficient,vandevondele2005quickstep,laino2005efficient,laino2006efficient,giese2016ambient,pederson2022dft}. 

An alternative approach, named the QM/MM-Ewald method and originally proposed in the context of semi-empirical QM treatments\cite{nam2005efficient,riccardi2005p,seabra2007implementation,walker2008implementation,holden2013periodic,kubavr2015new,nishizawa2016rapid,holden2019analytic,bonfrate2023efficient}, approximates the QM electron distribution by QM-atom-centered point charges. The long-range QM/MM electrostatics, i.e. the interaction between the reference cell QM density and  periodic MM charges outside of the reference cell, can then be handled using 
standard point-charge Ewald lattice sum methods, while the interactions between the QM density and nearby MM charges within the reference cell are computed exactly. 

The QM/MM-Ewald method has the benefit of simplicity and avoids the costly evaluation of the long-range electrostatic potential on a large number of grid points, but it assumes that the QM charge distribution is well approximated by point charges. This is a good approximation when there is a large separation between the reference cell QM charges and the MM charges in other cells from a large simulation box.
However, artifacts have been reported when applying this together with a DFT treatment of the QM region. In particular, difficulties were observed in converging the self-consistent field (SCF) iterations when using Mulliken QM charges and a diffuse basis\cite{holden2013periodic,holden2015erratum}, and a stable SCF required refitting the QM charges to the QM electrostatic potential during the SCF cycles\cite{holden2013periodic,holden2019analytic}.

\subsection{Multipole Approximation to Electrostatics}

Our approach to the long-range electrostatics starts from the simple and efficient QM/MM-Ewald method but extends the QM electronic description to a multipole representation\cite{laio2002hamiltonian,janowski2012ultrafast,alvarez2012asymptotic,giese2015multipolar,dziedzic2016tinktep,pan2018representation,olsen2019mimic,bolnykh2019extreme,dziedzic2019mutually,kirsch2021wavefunction,pan2021simplified,reinholdt2021fast,polonius2023lvc}. As we show later in the numerical results, this yields favorable convergence and SCF stability properties without the need to refit the QM charges during the SCF cycles.\cite{holden2013periodic,holden2019analytic}


To start, we partition the periodic QM/MM problem into two regions (Figure~\ref{fig:partitioning}). Region I contains the QM subsystem in the reference cell and the charges in the MM region that are close to the reference cell. Region II, the distant area, includes infinitely many atoms (i.e. the entire periodic lattice) and is where controlled approximations are made. It is assumed that the QM periodic images are always in region II, a convention maintained in our implementation and subsequent discussions. The electrostatic components of the energies $E_\mathrm{QM}$ and $E_\mathrm{QM-MM}$ are split into contributions from these two regions, labeled  $E^\I$ and $E^\II$. 

The QM-MM interaction within region I is computed using the exact QM density: 
\begin{equation} \label{eq:Eqm-mm1_pc}
    E_\mathrm{QM-MM}^\I = \sum_{i\in\I} \int \dd\mathbf{r} \frac{-\density q_i}{|\mathbf{r}-\RMM_i|}
    = \sum_{i\in\I} \sum_{\mu\nu} \int \dd\mathbf{r} \frac{-\phi_\mu(\mathbf{r})\phi_\nu(\mathbf{r})q_i}{|\mathbf{r}-\RMM_i|} \gamma_{\mu\nu}
\end{equation}
where we assume the MM charges are point charges $q_i$ at the MM atom positions, and we use $\gamma_{\mu\nu}$ to denote the one-body reduced density matrix element (1-RDM) in the QM region, and $\{\phi_\mu(\mathbf{r})\}$ to represent the atom-centered computational basis functions. (The extension to Gaussian-distributed MM charges is discussed later). 

For the QM-MM coupling in region II, 
we approximate the electrostatic potential around QM basis center $\RQM_\mu$, 
arising from an MM charge at position $\RMM_i$, using a multipole expansion (Taylor series) around $\RQM_\mu$,
($\alpha$ and $\beta$ index Cartesian components $x$, $y$ and $z$ in below):
\begin{align} \label{eq:taylor}
 \frac{1}{|\mathbf{r}-\RMM_i|} 
 & \approx \frac{1}{|\RQM_{\mu}-\RMM_i|} - \frac{(\RQM_\mu-\RMM_i)\cdot(\rr-\RQM_\mu)}{|\RQM_\mu-\RMM_i|^3} + \\
 & \frac{1}{2} \sum_{\alpha\beta} \frac{3(R^\mathrm{QM}_{\mu\alpha}-R^\mathrm{MM}_{i\alpha})(R^\mathrm{QM}_{\mu\beta}-R^\mathrm{MM}_{i\beta})-\delta_{\alpha\beta}|\RQM_\mu-\RMM_i|^2}{|\RQM_\mu-\RMM_i|^5}(r_\alpha-R^\mathrm{MM}_{i\alpha})(r_\beta-R^\mathrm{MM}_{i\beta}) \nonumber
\end{align}
For the QM density contribution of a basis pair $\mu$ and $\nu$ centered on different atoms, the expansion is performed at both centers and then arithmetically averaged. The Taylor series in Eq.~\ref{eq:taylor} has been truncated at the second order. The leading error in such an approximation is the missing third-order term that scales asymptotically as $1/|\RQM_\mu-\RMM_i|^4$. This is the first term in the series that decays sufficiently fast to ensure its infinite lattice sum (over the periodic MM images) absolutely converges in real space\cite {frenkel2023understanding}. This allows for control over the truncation error to achieve the desired accuracy by appropriately selecting Region I. We discuss how this can be accomplished in practice later.

It is straightforward to show that substituting the expansion (Eq.~\ref{eq:taylor}) into the QM-MM electrostatic integral leads to a distributed multipole expansion approximation centered on the QM atoms:
\begin{align} \label{eq:Eqm-mm2}
    E_\mathrm{QM-MM}^\II \approx \sum_{i\in\mathrm{MM(II)}}  \sum_{j\in\mathrm{QM(I)}}
    \Big[ Q_j q_i T_{ji} + \sum_\alpha \mu_{j\alpha} q_i T_{ji\alpha} + \sum_{\alpha\beta} \theta_{j\alpha\beta} q_i T_{ji\alpha\beta}
    \Big]
\end{align}
where $Q$, $\mathbf{\mu}$ and $\mathbf{\theta}$ are Mulliken atomic charges, dipoles and quadrupoles\cite{sokalski1983cumulative}, and the $T_{ji\cdots}$'s are the Taylor expansion coefficients, referred to as the multipole interaction tensors.
A similar multipole expansion approximation describes the interactions between the unit-cell QM and periodic QM images, yielding:
\begin{align} \label{eq:Eqm2}
    E_\mathrm{QM}^\II \approx \frac{1}{2}\sum_{i\in\mathrm{QM(II)}}\sum_{j\in\mathrm{QM(I)}} \Big[Q_jQ_iT_{ji} + \sum_\alpha \mu_{j\alpha}Q_iT_{ji\alpha} + \sum_{\alpha\beta}\theta_{j\alpha\beta}Q_iT_{ji\alpha\beta} 
    + 2\sum_{\alpha\beta} \mu_{j\alpha}\mu_{i\beta}T_{ji\alpha\beta} \Big]
\end{align}
We ignore the electronic exchange and correlation effects between QM images, due to their short-range nature. To manage the infinite lattice sum over region II, we utilize the identity $\sum_{i\in\II}=\sum_{i}-\sum_{i\in\I}$, converting the restricted sum over region II into an unrestricted sum over the whole lattice minus a finite real-space sum over region I. We treat the whole lattice sum by the Ewald method for multipoles\cite{aguado2003ewald} (see also SI), and we re-write the QM/MM energy of the periodic system as
\begin{align} \label{eq:qmmm_multipole}
    E(\Psi, & \RQM, \RMM)  =
    E_\mathrm{QM}^\mathrm{iso}(\Psi, \RQM) +  E_\mathrm{QM-MM}^\I(\density, \RQM, \RMM) \nonumber \\
&   -  \frac{1}{2}\sum_{ij\in\mathrm{QM(I)} \And i\neq j}  \Big[Q_jQ_iT_{ji} + \sum_\alpha \mu_{j\alpha}Q_iT_{ji\alpha} + \sum_{\alpha\beta}\theta_{j\alpha\beta}Q_iT_{ji\alpha\beta}
    + 2\sum_{\alpha\beta} \mu_{j\alpha}\mu_{i\beta}T_{ji\alpha\beta} \Big] \nonumber \\
&   + \frac{1}{2} \sum_{ij\in\mathrm{QM(I)}}  \Big[Q_jQ_i\psi_{ji} + \sum_\alpha \mu_{j\alpha}Q_i\psi_{ji\alpha} + \sum_{\alpha\beta}\theta_{j\alpha\beta}Q_i\psi_{ji\alpha\beta}
    + 2\sum_{\alpha\beta} \mu_{j\alpha}\mu_{i\beta}\psi_{ji\alpha\beta} \Big] \nonumber   \\
&   - \sum_{i\in\mathrm{MM(I)}}   \sum_{j\in\mathrm{QM(I)}}  
    \Big[ Q_j q_i T_{ji} + \sum_\alpha \mu_{j\alpha} q_i T_{ji\alpha} + \sum_{\alpha\beta} \theta_{j\alpha\beta} q_i T_{ji\alpha\beta} \Big] \\
&   + \sum_{i\in\mathrm{MM(unit~cell)}}   \sum_{j\in\mathrm{QM(I)}}
    \Big[ Q_j q_i \psi_{ji} + \sum_\alpha \mu_{j\alpha} q_i \psi_{ji\alpha} + \sum_{\alpha\beta} \theta_{j\alpha\beta} q_i \psi_{ji\alpha\beta} \Big] \nonumber \\
&  +E_\mathrm{MM}(\RQM, \RMM) \nonumber
\end{align}
where $E^\mathrm{iso}_\mathrm{QM}$ indicates the QM energy of an isolated QM subsystem i.e. without the periodic QM-QM electrostatic interactions. In Eq.~\ref{eq:qmmm_multipole}, the first two terms account for all the QM and QM-MM energies in the region I. The third and fourth terms sum to $E_\mathrm{QM}^\II$, and the fifth and sixth terms sum to $E_\mathrm{QM-MM}^\II$. The Ewald interaction tensors are denoted as $\psi_{ji\cdots}$ and are given in the SI. (When we write a summation range over the unit cell, we are assuming the minimum image convention).

In addition to the MM point charge model, it is useful to consider Gaussian smeared MM charges. Using such Gaussian charges reduces the so-called ``electron spill-out" effect\cite{nam2005efficient,laino2006efficient,brunk2015mixed}, where density that is very close to a point MM charge is overpolarized. In this context, $E_\mathrm{QM-MM}^\I$ is calculated as:
\begin{align}
    E_\mathrm{QM-MM}^\I 
    = \sum_{i\in\mathrm{MM(I)}} \sum_{\mu\nu} \int \dd\rr_1\dd\rr_2 \frac{-\phi_\mu(\rr_1)\phi_\nu(\rr_1)\phi_i(\rr_2)q_i}{|\mathbf{r}_1-\rr_2|} \gamma_{\mu\nu}    
    \label{eq:qmmmI}
\end{align}
where $\phi_i$ is a normalized Gaussian function with exponent $\kappa_i$, centered on MM atom $i$, representing its charge distribution. The energy expression for MM point charges (Eq.~\ref{eq:qmmm_multipole}) is then used with  a correction term to reconcile the discrepancy between Gaussian distributed charges and point charges in their interactions with QM multipoles, 
\begin{align} \label{eq:gauss_chrg_correction}
    \Delta E_\mathrm{QM-MM}^\II = -\sum_{i\in\mathrm{MM(II)}}\sum_{j\in\mathrm{QM(I)}}
    \Big[ Q_j q_i \hat{T}_{ji} + \sum_\alpha \mu_{j\alpha} q_i \hat{T}_{ji\alpha} + \sum_{\alpha\beta} \theta_{j\alpha\beta} q_i \hat{T}_{ji\alpha\beta}
    \Big]
\end{align}
where $\hat{T}_{ji\cdots}$ represents the multipole interaction tensors between point charges and Gaussian distributed charges (detailed in the SI). Since this correction decays exponentially fast with the charge separation, the infinite lattice sum over MM charges in the above expression reduces to a finite sum in real space, and the truncation can be made based on each MM charge exponent $\kappa_i$.

\subsection{Electronic Mean-Field Theory}

The Schr\"{o}dinger equation that $\Psi$ satisfies may be derived through the variational minimization of the QM/MM energy Eq.~\ref{eq:qmmm_multipole} with respect to $\Psi$. Within a mean-field approximation of the electronic structure, the wavefunction takes a single Slater determinant form, parameterized by the 1-RDM $\gamma$. Depending on the mean-field energy functional, the variational minimization of the QM/MM energy leads to the solution of Hartree-Fock (HF) or Kohn-Sham (KS) equations. The Fock/Kohn-Sham matrix that includes the QM/MM interactions is derived by taking the gradient of the QM/MM energy with respect to $\gamma$,
\begin{align}
   f_{\mu\nu} = f_{\mu\nu}^\I
   + \sum_{j\in\mathrm{QM(I)}}
 \Big[    v_j^\II \frac{\partial Q_j}{\partial \gamma_{\mu\nu}}
   + \sum_{\alpha} v_{j\alpha}^\II \frac{\partial \mu_{j\alpha}}{\partial \gamma_{\mu\nu}}
   + \sum_{\alpha\beta} v_{j\alpha\beta}^\II \frac{\partial \theta_{j\alpha\beta}}{\partial \gamma_{\mu\nu}}
   \Big]
\end{align}
In the above, $f^\I = \partial(E_\mathrm{QM}^\mathrm{iso}+E_\mathrm{QM-MM}^\I) /{\partial \gamma}$ is the Fock operator for the non-periodic QM/MM system consisting of the unit-cell QM atoms and the region I MM atoms. The potential $v^\II$ that the QM multipoles experience due to their interactions with region II is derived by differentiating the multipole-dependent terms in Eq.~\ref{eq:qmmm_multipole} (and Eq.~\ref{eq:gauss_chrg_correction} if using Gaussian-distributed MM charges), specifically, $(v_j^\II,v_{j\alpha}^\II,v_{j\alpha\beta}^\II)=\partial E^\II/\partial (Q_j,\mu_{j\alpha},\theta_{j\alpha\beta})$. The response of the Mulliken multipoles to changes in the 1-RDM is given by overlap-type integrals,\:
\begin{align}
&    \frac{\partial Q_j}{\partial \gamma_{\mu\nu}} = -\frac{1}{2} (\delta_{\mu\in j}+\delta_{\nu\in j}) \langle\mu|\nu\rangle \\
&    \frac{\partial \mu_{j\alpha}}{\partial \gamma_{\mu\nu}} = -\frac{1}{2} (\delta_{\mu\in j}+\delta_{\nu\in j}) \langle\mu|(r_\alpha-R_{j\alpha})|\nu\rangle\\
&    \frac{\partial \theta_{j\alpha\beta}}{\partial \gamma_{\mu\nu}} = -\frac{1}{2} (\delta_{\mu\in j}+\delta_{\nu\in j}) \langle\mu|(r_\alpha-R_{j\alpha})(r_\beta-R_{j\beta})|\nu\rangle
\end{align}
where $\delta_{\mu\in j}$ signifies whether orbital $\mu$ is centered on atom $j$, and $\langle\mu|f(\rr)|\nu\rangle$ is a shorthand notation for $\int \dd\rr \phi_\mu(\rr)f(\rr)\phi_\nu(\rr)$.
As discussed earlier, the Taylor expansion (Eq.~\ref{eq:taylor}) is conducted at both centers of orbitals $\mu$ and $\nu$ and then averaged, and thus the $1/2$ factor arises. Such an averaged expansion is equivalent to symmetrizing the Fock operator resulting from a single-sided Taylor expansion centered on only one orbital.

The nuclear gradient required to perform MD is obtained by differentiating the QM/MM energy with respect to the atom positions. The nuclear gradient of the interaction energy in the region I ($E_\mathrm{QM}^\mathrm{iso}+E_\mathrm{QM-MM}^\I$) is well-documented in the literature for non-periodic QM/MM systems (see e.g., Ref.~\cite{gao1996methods}). The $E^\II$ gradient contains both the response of the multipole interaction tensors $T_{ji\cdots}$, $\psi_{ji\cdots}$ and $\hat{T}_{ji\cdots}$, and of the multipole moments $Q$, $\mu$ and $\theta$. The interaction tensors, as purely geometric functions, are independent of the multipole moments, and it is straightforward to differentiate them with respect to the atom positions (see SI for detailed expressions). The multipole response formally depends on both a Pulay term (response due to the basis center movement) and the 1-RDM response. However, the 1-RDM response is not explicitly needed if the wavefunction is at a variational stationary point, i.e. as long as the self-consistent HF/KS equations are converged. Given these considerations, the $E^\II$ gradient with respect to a QM atom $j$ (denoted $\nabla_j E^\II$) can be written as
\begin{align}
    \nabla_{j} E^\II & = \sum_{\mu\nu} \gamma_{\mu\nu} (\delta_{\nu\in j}-\delta_{\mu\in j}) \times \nonumber \\
    \Big[
&   v_\mu^\II   \langle\nabla \nu|\mu\rangle 
  + \sum_\alpha v_{\mu\alpha}^\II \langle\nabla\nu|(r_\alpha-R_{\mu\alpha})|\mu\rangle 
  + \sum_{\alpha\beta} v_{\mu\alpha\beta}^\II \langle\nabla\nu|(r_\alpha-R_{\mu\alpha})(r_\beta-R_{\mu\beta})|\mu\rangle 
    \Big]
\end{align}
where on the right hand side $\nabla$ is the gradient with respect to the electron coordinate, and $v_{\mu\cdots}^\II$ is the potential experienced by the atomic multipoles that the orbital $\mu$ is centered on.
To derive the above expression, we have assumed that the 1-RDM is symmetric, and also used the $\nabla_j\phi_\mu(\rr) = -\delta_{\mu\in j} \nabla\phi_\mu(\rr)$ property for an atom-centered orbital, and the identity $\langle\nabla \nu| f(\rr) | \mu\rangle + \langle\nu|\nabla f(\rr)|\mu\rangle + \langle\nu|f(\rr)|\nabla\mu\rangle = \mathbf{0}$. Finally we note that the $E^\II$ gradient with respect to an MM atom only involves the gradient of multipole interaction tensors as there are no associated Pulay contributions.

\subsection{Pseudo-bond Approach for the QM/MM boundary}

When there are covalent bonds between some of the QM atoms and MM atoms, such as in the case of a QM treatment of amino acid sidechains with an MM treatment of the backbone, dangling bonds arise in the QM subsystem from the QM/MM partitioning. Usually, the dangling bonds are either capped by an atom\cite{antes1999adjusted,zhang1999pseudobond,dilabio2002simple,zhang2005improved,von2005variational,slavivcek2006multicentered,shao2007yinyang,xiao2007design,parks2008pseudobond} or by an orbital\cite{thery1994quantum,gao1998generalized,philipp1999mixed,sun2014exact}. We adopt the so-called pseudo-bond approach here because of the minimal changes it requires in a QM/MM implementation\cite{zhang1999pseudobond,zhang2005improved,parks2008pseudobond}. The method introduces a specially parameterized cap atom to the dangling bond, placed at the location of the MM atom that is supposed to form the bond with QM atoms. For instance, in the above sidechain-backbone partitioning, a special fluorine atom carrying special basis sets and a pseudo-potential is positioned where the backbone $\alpha$-carbon would be. Consequently, the C$_\alpha$ atom is treated as having dual QM and MM identities -- it forms covalent bonds with the QM sidechain via quantum mechanical interactions and also interacts with the MM backbone via a force field. The fluorine basis and pseudo-potential are parameterized to mimic an $sp^3$ hybridized C$_\alpha$, bonded to the sidechain group \ch{-R}. This is achieved by fitting the parameters to reproduce the geometry, charge distribution (excluding the methyl moiety), and relative protonation/phosphorylation energetics of a \ch{R-CH3} molecule by a \ch{R-F} molecule. Herein, we use the same basis and pseudo-potential forms as in Refs.~\cite{zhang2005improved,parks2008pseudobond} and optimize for three DFT/basis combinations, $\omega$B97X-V\cite{mardirossian2014omegab97x}/6-31G**, $\omega$B97X-V/6-311G** and $\omega$B97X-3c\cite{muller2023omegab97x}. The parameterization details and the resulting parameters are provided in the SI.

\subsection{Implementation}
We have implemented our QM/MM-Multipole method within the \textsc{PySCF} package taking advantage of GPU acceleration via the \textsc{GPU4PySCF} module. Given the full system atom positions, the QM atom element types, and MM charges (and optionally radii), \textsc{PySCF} is tasked with computing $E_\mathrm{QM}$, $E_\mathrm{QM-MM}$, and their nuclear gradients with respect to all atoms. The computationally expensive SCF calculations are offloaded to GPU routines in \textsc{GPU4PySCF}. 
The QM-MM electrostatic integral within region I (Eq.~\ref{eq:qmmmI}) is computed using the GPU implementation of the three-center-two-electron integrals in \textsc{GPU4PySCF}. In addition, the interaction tensors $T$, $\hat{T}$, and $\psi$, are computed on the GPU through CuPy\cite{cupy_learningsys2017} array operations, while the tensor contractions 
between the interaction tensors and multipoles are also performed by CuPy.
In our QM/MM implementation, it is not necessary to employ a particle-mesh variant of Ewald, which uses a real-space grid to interpolate the charges, to improve the computational scaling. This is because by choosing the Ewald real-space cutoff to be roughly the box length, the computational cost of both the real-space and $k$-space sums scales linearly with the number of MM atoms since we only compute the QM-QM and QM-MM Ewald energies in \textsc{GPU4PySCF}, and not the MM-MM Ewald energy. Aside from the overlap-like one-electron integrals, used for computing multipole moments and their gradients, which are evaluated on the CPU due to their relatively low cost, all QM/MM-related computational tasks are accelerated by the GPU. 

We use LAMMPS\cite{LAMMPS} to compute $E_\mathrm{MM}$ and its nuclear gradients. This is straightforwardly achieved within LAMMPS input files, by turning off the bonded interactions within the QM subsystem and the electrostatics between the QM and MM atoms. In our current setup, \textsc{GPU4PySCF} and LAMMPS run concurrently in the background and communicate with an MD integrator, i-PI\cite{kapil2019pi}. At each MD step, i-PI sends the box dimensions and atom positions via internet or Unix sockets to both \textsc{GPU4PySCF} and LAMMPS and collects the computed energy and forces from them to integrate the nuclear dynamics. Such a communication scheme allows for the overlap of the LAMMPS CPU computation with GPU computation by \textsc{GPU4PySCF}. We note that LAMMPS itself can serve as the MD integrator as well if one desires to eliminate a dependency on i-PI. Presently since there is no communication between \textsc{GPU4PySCF} and LAMMPS, any choice of MD code can be easily adopted into this QM/MM communication scheme and requires no modification if it comes with an out-of-the-box i-PI interface, or if not,  only small changes are needed to enable socket communication with i-PI.

\section{Computational Details}

\subsection{Common Setup}
We outline here computational details that the simulations below share in common.  In all the simulations, we modeled the MM charges as Gaussian-distributed charges when computing the QM-MM interactions, with the exponent of the Gaussian on an atom given by the square of the inverse of its covalent radius in Ref.~\cite{pyykko2009molecular}. The MM partial charges (used in $E_\text{QM-MM}$) and the Lennard-Jones parameters between the QM and MM atoms (used in $E_\text{MM}$) were kept the same as in the classical force field used in the simulation. \red{We adopted an Ewald real-space cutoff roughly the size of the box length and estimated the Ewald parameters according to Ref.~\cite{lindbo2011spectral} given a desired $10^{-8}$ Hartree accuracy.}
The density fitting approximation to the electron repulsion integrals was employed, using the def2-SVP-JKFIT basis as the auxiliary fitting basis for the def2-SVPD basis, and the 6-311G**-RIFIT basis\cite{tanaka2013optimization} for all the other atomic bases. For the pseudo-bond atoms, the fitting basis was generated by \textsc{PySCF}\cite{pyscf} using an even-tempered Gaussian expansion. 
The SCF convergence criteria were chosen to be $10^{-10}$ Hartrees for the energy and $10^{-6}$ Hartrees for the orbital gradient. In the MD simulations, we used the ``time-reversible always stable predictor-corrector'' method (ASPC)\cite{kolafa2004time,kuhne2007efficient} to predict the 1-RDM for the current MD step from the four preceding MD steps. The predicted 1-RDM was used as the initial guess for SCF and for dynamically determining region I based on a given error threshold (detailed further below). In all QM/MM $NVT$ simulations, the temperature was controlled by a Langevin thermostat with a damping time constant of 100 fs. A time step of 0.5 fs was used for all QM/MM MD unless stated otherwise.

\subsection{QM Water in MM Water}

For this application, we considered three QM-water-in-MM-water systems, consisting of 7 QM water molecules embedded in respectively 1, 221, and 33234 SPC/Fw\cite{wu2006flexible} MM water molecules. 
The initial configuration of the 221 and 33234 MM water systems was generated using Packmol\cite{martinez2009packmol} by placing 228/33241 water molecules in cubic boxes with 
19.002~\AA~and 100.013~\AA~side lengths, according to a 0.99403 kg/L water density.
The system was energy minimized and equilibrated for 50 ps at 298.15 K at the SPC/Fw MM level. The temperature was controlled by a Nose-Hoover chain and the dynamics were integrated using a 0.5 fs time step. The final equilibrated configuration was used for subsequent QM/MM simulations, where the 7 closest waters to the box center were selected as QM waters. 
All QM/MM calculations employed the PBE\cite{perdew1996generalized}/def2-SVPD level of theory. The 7-QM-water-in-1-MM-water system was built by extracting the QM water configuration from the equilibrated 228-water structure and placing it at the center of a cubic box with a 27.399~\AA~side length. An MM water molecule was then positioned along one of the box diagonals with varying distances from the box center.

\subsection{Guanosine Triphosphate Hydrolysis in Microtubules}
A model of guanosine triphosphate (GTP) hydrolysis in microtubules (MTs) was taken from a previous study\cite{beckett2023unveiling} by Beckett et al. The system consists of a GTP-bound compacted inter-dimer of tubulin solvated in 56306 water molecules with 149 \ch{Na+} and 112 \ch{Cl-} ions in a cubic box with a 121.5~\AA~side length. The previously equilibrated configuration was directly used as the starting point for our QM/MM simulation. The QM treatment was applied to the tri-phosphate part of the $\beta$-site GTP molecule, the coordinated \ch{Mg^{2+}} ion, specific residue sidechains from both the $\alpha$ and $\beta$ subunits ($\alpha$:Arg2, $\alpha$:Glu254, $\alpha$:Lys352, $\beta$:Asp67, $\beta$:Glu69, $\beta$:Asn99 and $\beta$:Thr143) and 19 solvation waters. The rest of the system was described by the CHARMM36m force field\cite{huang2017charmm36m}. The pseudo-bond approach was used for the broken C$_\alpha$\ch{-C}$_\beta$ bonds of the amino acids and for the broken C4'$-$C5' bond for the GTP molecule. The backbone atoms of the QM residues and the H4', O4', C1', H1', C3', H3', O3', and H3T (CHARMM naming) atoms of the GTP molecule were set to have zero charge
when computing the QM-MM coupling. All the MT simulations were performed in the $NVT$ ensemble at 310 K, and the QM region was described at the PBE\cite{perdew1996generalized}/6-311G**, \red{B3LYP\cite{stephens1994ab}/6-311G*,} $\omega$B97X-V\cite{mardirossian2014omegab97x}/6-311G** or $\omega$B97X-3c\cite{muller2023omegab97x} levels of theory.

\subsection{Chorismate Mutase}
The chorismate mutase (CM) model we used is based on a previous study\cite{ray} by Ray et al. The system comprises a chorismate-bound enzyme solvated in 14821 water molecules with 12 \ch{Na+} ions in a box measuring 79.006~\AA$\times$79.682~\AA$\times$79.030~\AA. We followed the same simulation setup as Ray et al., except for using our QM/MM-Multipole scheme and varying the QM region definition and DFT functional. 

We considered two QM region definitions: one including the substrate and the catalytic residue Arg90 (S+R90), and another including the substrate with Arg90, Arg7, and Glu78 (S+R90+R7+E78). The pseudo-bond approach was used for the broken C$_\alpha$\ch{-C}$_\beta$ bonds of the QM residues, and the backbone atoms of the QM residues were set to have zero charge when computing the QM-MM coupling.

In the timing tests, the QM region was described by the PBE/6-311G**, $\omega$B97X-V/6-311G** or $\omega$B97X-3c levels of theory. In the energy conservation test, we tested $\omega$B97X-3c with the S+R90 region. In the production runs for the catalytic kinetics, the QM region was described by PBE/6-31G**, $\omega$B97X-3c, or a custom refined version of $\omega$B97X-3c (see below for a detailed description of the customization).

We equilibrated the structure from Ray et al. using our QM(S+R90)/MM MD for 12 ps at 300 K, with a 1 fs time step, at both the PBE/6-31G** and $\omega$B97X-3c levels of theory. The last frame of the PBE equilibration was used to initiate a 15-ps-long steered MD (SMD), biasing several inter-atomic distances to flip the Arg90 residue to form a hydrogen bond between the chorismate enolpyruvyl carboxylic oxygen and the arginine $\epsilon$-nitrogen. The SMD was performed at the PBE/6-31G** level, and the detailed protocol can be found in the SI. The resulting structure with a flipped Arg90 forms two hydrogen bonds (HBs) with the chorismate while the structure taken from Ref.\cite{ray} features one chorismate-Arg90 hydrogen bond. 

Additional QM/MM equilibration at both the PBE/6-31G** and $\omega$B97X-3c levels was started from the last frame of the SMD, and run for 5 ps and 12 ps respectively. The resulting structures were further equilibrated for 12 ps at the same level of theory with a larger QM region (S+R90+R7+E78). The $\omega$B97X-3c equilibration was followed by an additional 12 ps of equilibration at the refined $\omega$B97X-3c level.

For each combination of Arg90 conformation, QM region definition, and level of quantum theory, 11 configurations were selected from the last 10 ps of equilibration to initiate on-the-fly probability enhanced sampling (OPES) flooding simulations\cite{ray2022rare}. The configurations after the first 2-ps of each equilibration were used for initiating an SMD simulation to estimate the reaction free energy profile.
The reaction coordinate (RC) $\xi$ for both OPES flooding and SMD was the bond-making C1$-$C9 distance minus the bond-breaking C3$-$ester oxygen distance. A harmonic restraint with a 1000 kJ/mol/\AA$^2$ force constant was used to drive the RC from its initial reactant value ($\xi\approx$1.8~\AA) to -0.9~\AA~over 7.5 ps of SMD. The free energy surface obtained from the SMD was used to determine the barrier height parameter and the bias-excluded region defined as $\xi<\xi^\mathrm{excl}$ for the following OPES flooding. Detailed OPES flooding protocols are provided in the SI. All the enhanced sampling simulations (SMD and OPES flooding) used a 1 fs time step, with the enhanced sampling functionality provided by the PLUMED library\cite{tribello2014plumed,bonomi2019promoting}.

We also performed geometry optimization and nudged elastic band (NEB) simulations to find reaction transition states starting from both the 1HB and 2HB Arg90 conformations. The first frames of the $\omega$B97X-3c SMD with 1HB/2HB Arg90 and the S+R90 QM region were energy minimized at the same level of theory to produce local minimum energy reactant configurations. 
The last frames of the $\omega$B97X-3c SMD were geometrically aligned to the reactant structures, and their product-like active-site conformations were combined with the geometries in the reactant state for the remaining part of the system. 
The combined conformations were energy minimized to find the product state minima. Eight images were interpolated between the resulting reactant and product states using the image dependent pair potential method\cite{smidstrup2014improved}. NEB optimization was performed without a climbing image, targeting a maximum force of 0.15 eV/\AA, and then with a climbing image using the same convergence criterion. 
The geometry minimization and NEB optimization were performed on the QM/MM PES with the S+R90 QM region and the $\omega$B97X-3c functional, with the atoms of S+R90+R7+E78 allowed to move. All energy minimization was performed using i-PI and all NEB calculations were carried out using the Atomic Simulation Environment (ASE)\cite{larsen2017atomic}.

We extracted the substrate and Arg90 geometries from the NEB conformations and performed single-point energy calculations in the gas phase. \red{The Arg90 sidechain was capped by a hydrogen positioned along the C$_\alpha$-C$_\beta$ direction at a distance of 1.08~\AA~from C$_\beta$.} The energies were computed using PBE/6-31G** and $\omega$B97X-3c on every NEB image, and LNO-CCSD(T)\cite{rolik2011general} at five geometries: the reactant, the product, the transition state, and the two adjacent NEB images to the transition state. \red{We performed the LNO-CCSD(T) calculations with cc-pVTZ and cc-pVQZ basis sets and used the two-point extrapolation formula\cite{neese2011revisiting} to extrapolate both the HF and correlation energies to the complete basis set (CBS) limit. All LNO calculations were performed with our recent implementation\cite{zhang2024performant} in \textsc{PySCFAD}\cite{zhang2022differentiable}. A natural occupation threshold of $10^{-5}$ was chosen for truncating the local virtual natural orbitals, while a natural occupation threshold of $10^{-4}$ was used for truncating the occupied natural orbitals. Such a choice of thresholds was found to give 0.05 kcal/mol mean absolute error in the LNO-CCSD(T)/cc-pVDZ energy on the five 1HB NEB geometries and 0.02 kcal/mol on the 2HB NEB geometries, compared to the canonical CCSD(T)/cc-pVDZ results. All the cc-pVDZ, cc-pVTZ, and cc-pVQZ calculations were also found to be converged to within 0.06 kcal/mol in MAE with the $10^{-4}$ occupied/$10^{-5}$ virtual occupancy thresholds when comparing to the energies with larger thresholds ($2\times 10^{-4}$ occupied/$2\times10^{-5}$ virtual).}  To reduce the error of the underlying quantum theory used in our MD simulations, we then refined the exchange parameters of the $\omega$B97X-3c functional to reproduce the \red{LNO-CCSD(T)/CBS} reaction energy barriers in both the 1HB and 2HB Arg90 conformations, optimizing both the fraction of short-range and of long-range exchange to exactly reproduce the two reaction barriers.
We then checked to make sure that the resulting refined $\omega$B97X-3c functional also improved the energetics on the non-training points as well. We observed more accurate energies and curvature around the transition state, noting that these are related to the probability of successful barrier crossing in classical rate theory\cite{hanggi1990reaction}.

\section{Results and Discussion}
\subsection{Error Control via Charge-Octupole Estimation}
\label{subsec:rcut}

\begin{figure}
    \centering
    \includegraphics[width=15cm]{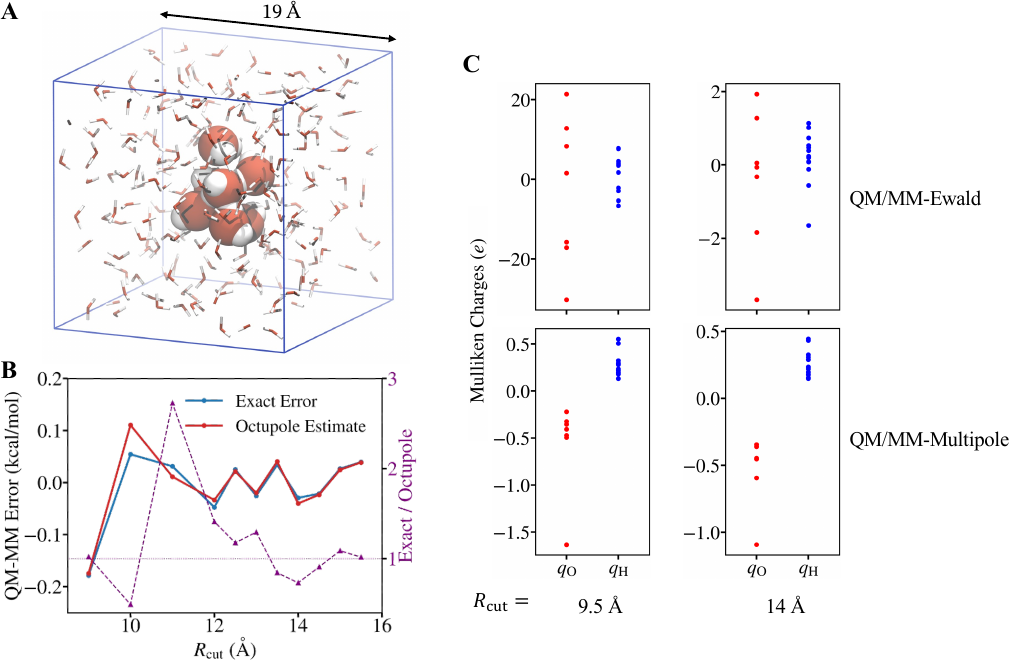}
    \caption{QM/MM analysis for a QM-water-in-MM-water system. (A) The configuration of the system where QM/MM calculations were performed. The QM water molecules are shown as spheres and the MM water molecules are shown as sticks. (B) The error of QM/MM-Multipole (up-to-quadrupoles) QM-MM energy as a function of the \red{region I} cutoff radius $R_\mathrm{cut}$. (C) Mulliken charges of the QM atoms after up to 200 cycles of SCF with QM/MM-Ewald or QM/MM-Multipole using two $R_\mathrm{cut}$ values.}
    \label{fig:228wat}
\end{figure}

As discussed earlier, we truncate the multipole expansion 
at the quadrupole order (second order in the Taylor expansion) \red{for the electrostatics between region II and the QM density in the reference cell}, 
and thus the leading error in our approximation is the missing charge-octupole (the third order) interaction.
This suggests that the interaction energy between the region II charges and the QM octupoles can serve as an error estimate,
\begin{align}
    \mathrm{Err}_\mathrm{QM-MM}\approx \sum_{i\in\mathrm{MM(II)}}\sum_{j\in\mathrm{QM(I)}} \sum_{\alpha\beta\gamma} \Omega_{i\alpha\beta\gamma}q_iT_{ji\alpha\beta\gamma}
\end{align}
This estimate assumes that region I is sufficiently large that any higher-order terms (fourth order and above) are negligible.
From here on, we consider region I to be defined by a sphere of radius $R_\mathrm{cut}$ centered at the QM subsystem geometric center. \red{This cutoff truncates the short-range interactions of the QM octupoles and higher order multipoles with the charges outside region I, and should not be confused with the Ewald real-space cutoff.}

To validate the charge-octupole (octupole for short) error estimate, we examined a simple QM-water-in-MM-water system where computing the exact QM-MM electrostatic coupling energy is feasible. The system consists of 7 QM water molecules embedded in 221 MM water molecules (Figure~\ref{fig:228wat}A). The exact QM-MM electrostatic coupling was computed by the standard Ewald method, treating the electron density on the DFT grid as one group of point charges and the MM water molecules as another. The electron density used in the test was obtained from a converged QM/MM-Multipole SCF with $R_\mathrm{cut}=$15.5~\AA. In Figure~\ref{fig:228wat}B, we show the exact error of including up to quadrupoles ($E_\mathrm{exact}-E_\mathrm{approx}$) versus the region II octupole error estimate as a function of $R_\mathrm{cut}$. (We did not re-optimize the density with different $R_\mathrm{cut}$ in this test, thus the $R_\mathrm{cut}$ dependency is entirely due to the varying number of exactly computed QM-MM interactions). We find a good agreement between the octupole estimate and the exact error beyond $R_\mathrm{cut}\approx12.5$~\AA. For a smaller $R_\mathrm{cut}$, the octupole error estimate remains qualitatively correct and is of the same order of magnitude as the exact error.

\begin{figure}
    \centering
    \includegraphics[width=15cm]{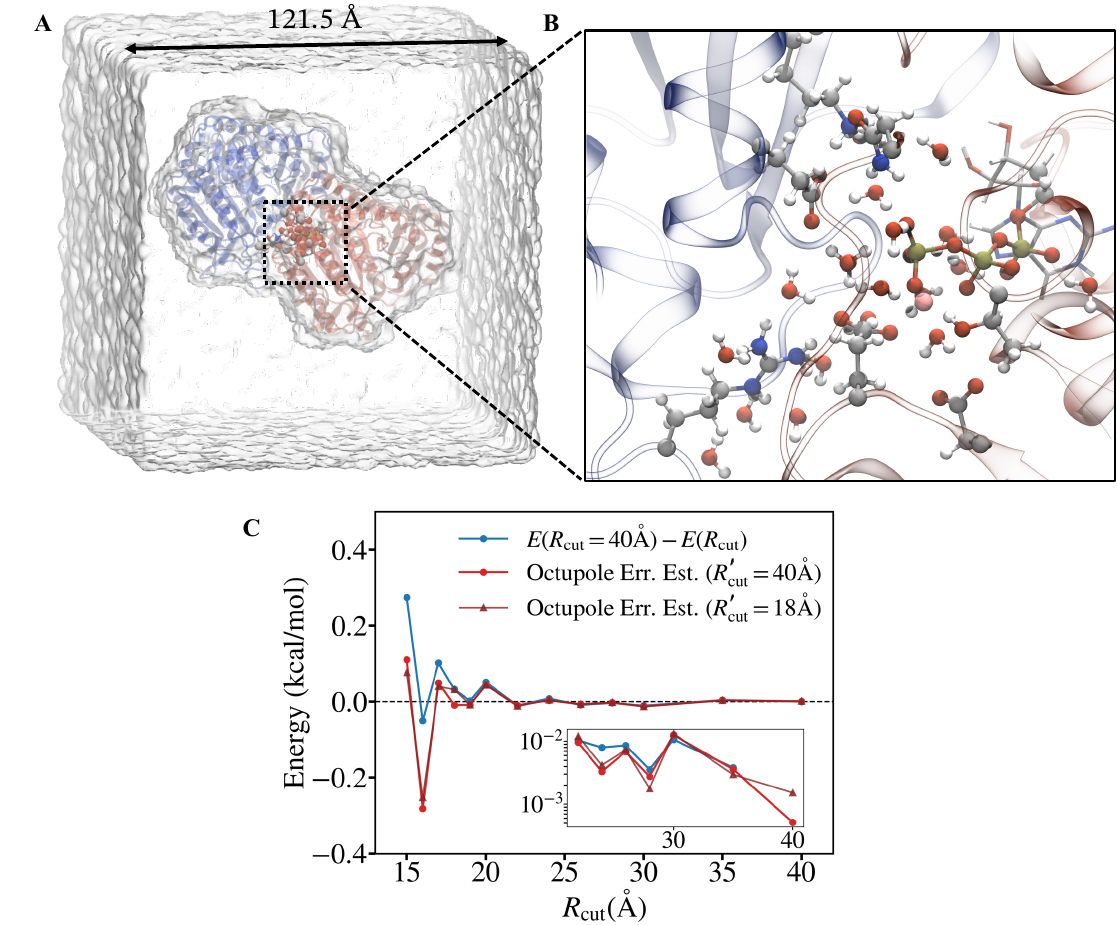}
    \caption{QM/MM analysis of a model for microtubule-mediated GTP hydrolysis. (A) The system configuration used for the QM/MM calculations. The QM atoms are shown as spheres, the protein is depicted as ribbons (with subunit $\alpha$ in blue and subunit $\beta$ in red), and the rest of the system is represented by transparent surfaces. (B) A zoomed-in view of the QM region. The QM atoms are shown as balls and sticks, the MM part of GTP is represented by thin sticks, and the protein is shown in transparent blue ($\alpha$ subunit) and red ($\beta$ subunit). The atoms are colored as follows: hydrogen (white), oxygen (red), nitrogen (blue), carbon (gray), magnesium (pink), and phosphorus (gold). (C) The error in the QM/MM-Multipole SCF-converged energy as a function of the \red{region I} cutoff radius $R_\text{cut}$, assuming the energy obtained with $R_\text{cut} = 40$ Å is the ground truth. The octupole error estimate was computed using the density converged with a cutoff of $R'_\text{cut}$.}
    \label{fig:mt}
\end{figure}

We next examine a more complex system, a model of GTP hydrolysis in microtubules. The system composition and QM subsystem definition are detailed in the Computational Details section as well as shown in Figure~\ref{fig:mt}A and Figure~\ref{fig:mt}B. The electronic structure was described at the $\omega$B97X-3c level in this analysis. We estimated the exact error by comparing the energy at $R_\mathrm{cut}=40$~\AA~(which we take as exact) with the SCF-converged energy $E(R_\mathrm{cut})$ at various $R_\mathrm{cut}$ values. Unlike in the previous example, this error depends on $R_\mathrm{cut}$ not only through the omission of octupole and higher interactions with region II but also through the $R_\mathrm{cut}$-dependency in the converged QM density, from the absence of polarization effects at the above-quadrupole level by region II charges. According to the linear response theory of HF/KS\cite{yamaguchi1994new}, the error in the density is proportional to the error in the electronic Hamiltonian. As such, the lack of octupole polarization by a charge at distance $R$ introduces a density error that scales in the same way as the charge-octupole interaction ($\propto 1/R^4$). Integrating over the whole region II, this gives an accumulated error scaling as $\int_{R_\mathrm{cut}}^\infty 4\pi R^2 /R^4 \dd R\propto 1/R_{\mathrm{cut}}$. Due to the variational condition, the total energy error scales quadratically with the density error, yielding a $1/R_\mathrm{cut}^2$ decay, faster than the $1/R_\mathrm{cut}$ scaling of the total octupole interaction with region II. Hence, the leading term of the error at large $R_\mathrm{cut}$ is still the missing octupole interaction, the same as in the previous test with a frozen density. As shown in Figure~\ref{fig:mt}C, the octupole error estimate obtained by simply summing the missing charge-octupole interactions agrees well with the ``exact" error at sufficiently large $R_\mathrm{cut}$. Furthermore, since the energy error due to the density error decays quickly ($1/R_\mathrm{cut}^2$), the octupole error estimate may be computed using a density converged with a small region I (defined by $R_\mathrm{cut}'$), and we show this is indeed the case in Figure~\ref{fig:mt}.

We now discuss the practical use of such an octupole error estimate to determine $R_\mathrm{cut}$ in production simulations. At the first MD step, one may perform a QM/MM-Multipole SCF calculation with an empirically chosen $R_\mathrm{cut}$ to generate a QM density guess. A new $R_\mathrm{cut}$ can be determined by incrementing $R_\mathrm{cut}$ from a chosen lower bound until the octupole error estimate computed with the density guess falls below a desired accuracy $\epsilon$. If the new $R_\mathrm{cut}$ is larger than the initially guessed one, one may either rerun the first MD step with the determined $R_\mathrm{cut}$ or simply discard this step from the statistics. In the following MD steps, a proper $R_\mathrm{cut}$ can be estimated on the fly using the ASPC-predicted density from previous steps.

\subsection{Timings}

We evaluated the computational efficiency of our QM/MM-Multipole approach on two biomolecular systems: a chorismate mutase model and a model for microtubule-mediated GTP hydrolysis. We note that many aspects of these calculations (ranging from the auxiliary basis, here chosen as 6-311G-RIFIT in all cases below, to the detailed implementation)  can be further tuned to improve the efficiency. Thus the timings reported below should be viewed only as an initial guide to the performance.


For the PBE \red{and B3LYP} calculations, the pseudo-bond parameters for $\omega$B97X-V were used to avoid building another parametrization. The wall time per MD step was averaged over 100 MD steps, with a fixed $R_\mathrm{cut}=25$\AA. 
The MT system contains 156 QM atoms, and the CM system has 43 QM atoms in the S+R90 definition of the QM subsystem and 72 QM atoms in the S+R90+R7+E78 definition of the QM subsystem. The number of computational basis functions ($N_\mathrm{AO}$) is 1756, 532, and 846 using the 6-311G** basis in the MT and for the two CM QM subsystem definitions, respectively\red{, and $N_\mathrm{AO}$=1513, 475, and 744 for the 6-311G* basis set}. The $\omega$B97X-3c functional comes with a specially parameterized double-zeta basis, and $N_\mathrm{AO}=1309$, 398 and 637 in the MT and two CM QM subsystem definitions.
The timing results are provided in Table~\ref{tab:timing}. 

\begin{table}[h]
    \centering
    \begin{tabular}{llllll}
    \hline
System     & PBE/6-311G** & \red{B3LYP/6-311G*} & $\omega$B97X-3c & $\omega$B97X-V/6-311G** \\
    \hline
MT      & 1900$\pm$90 & \red{1510$\pm$70} & 730$\pm$20      & 370$\pm$20 \\
CM (S+R90) & 11200$\pm$600 & \red{9900$\pm$500} & 6600$\pm$300     & 4400$\pm$200       \\
CM (S+R90+R7+E78) & 6900$\pm$300 & \red{5800$\pm$300} & 3290$\pm$80     & 1790$\pm$80       \\
\hline                
    \end{tabular}
    \caption{Computational performance of the QM/MM MD simulations using 32 CPU cores (AMD EPYC 7742) and one A100 GPU. The reported values are the mean number of steps per day, with the fluctuations computed from the standard deviation over 100 MD steps.}
    \label{tab:timing}
\end{table}

We observe that in our current implementation of periodic ab initio QM/MM MD, simulations with a hybrid density functional ($\omega$B97X-3c) on systems with $> 100$ atoms ($>1000$ atomic basis functions) can be run with a throughput of $\sim$1000 MD steps per day on a single GPU. Although a careful comparison will require controlling many other details and will be considered in a later study, \red{the speed of B3LYP/6-311G* on the MT system} appears to be roughly comparable to that reported \red{for B3LYP on a system with a similar size} in recent GPU implementations of non-periodic ab initio QM/MM MD\cite{cruzeiro2023terachem}.


\subsection{SCF Stability}

If one ignores the dipole and quadrupole terms, the QM/MM-Multipole approach reduces to the QM/MM-Ewald approach using Mulliken charges\cite{nam2005efficient}. \red{(More precisely, the original QM/MM-Ewald approach additionally fixes region I to be the unit cell but we allow a flexible choice of region I herein).} It is known that the QM/MM-Ewald method with Mulliken charges suffers from instabilities in the SCF optimization when diffuse basis functions are used. The instability is related to the definition of the Mulliken charges, which becomes increasingly unsatisfactory in diffuse bases (although a description of the electrostatic potential at long distances is still provided by the full set of Mulliken multipoles).
In the test system shown in Figure~\ref{fig:228wat}A, we show that we could not converge the SCF within 200 cycles using the QM/MM-Ewald approach even when $R_\mathrm{cut}$ was set to half the box length. After extending region I beyond the unit cell, with $R_\mathrm{cut}=$14~\AA, we were able to finally converge the QM/MM-Ewald SCF.
The converged electron density, however, still has large errors revealed by large positive/negative Mulliken charges on the oxygen/hydrogen atoms. In contrast, the QM/MM-Multipole approach is always stable and produces reasonable Mulliken charges, even with the smaller $R_\mathrm{cut}$.

\subsection{Energy Conservation}
\begin{figure}
    \centering
    \includegraphics[width=8cm]{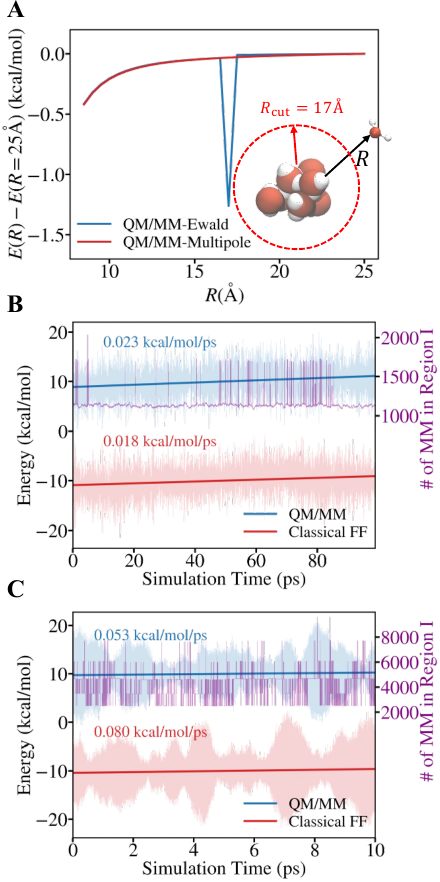}
    \caption{Energy conservation of the QM/MM-Multipole method. (A) The potential energy surface with a constant \red{region I} $R_\mathrm{cut}$ as a function of an MM water (shown as balls and sticks) distance from the QM cluster (shown as spheres) center. (B) Energy conservation in a QM-water-in-MM-water system. Linear fits to the energy time series are plotted as solid lines, and the text in the plots shows the slopes of the fits. (C) Energy conservation in a chorismate mutase model.}
    \label{fig:econsv}
\end{figure}

Since we treat the electrostatics in regions I and II differently, the potential energy surface of our QM/MM-Multipole approach is not strictly smooth. In Figure~\ref{fig:econsv}A, we check the smoothness of the energy surface when one MM water is moved from region I to region II with a fixed $R_\mathrm{cut}=17$~\AA. In contrast to the large energy jumps observed for QM/MM-Ewald, we see smooth (on the scale of the plot) energies in the QM/MM-Multipole approach. The large jump of the QM/MM-Ewald energy right at $R=R_\mathrm{cut}$ occurs when the boundary cuts through the MM water molecule. This is because although the leading electrostatic interaction with a water molecule should only involve the water dipole, when the water molecule is cut, the electrostatics in QM/MM Ewald is separated into interactions with individual charges on the water molecule, some of which are treated exactly and some of which are treated in the QM monopole approximation. There is thus a large residual charge interaction term which is unphysical.
The QM/MM Ewald energy still has a 
noticeable jump, however, even when we compare energies just before and after a whole water molecule is moved to region II, when this charge interaction error is absent.

We next examine whether the QM/MM-Multipole PES is smooth enough not to lead to a significant energy drift during long MD runs. We first tested this on a system consisting of 7 QM water molecules and 33234 MM water molecules. A harmonic potential with a force constant of 0.2 kcal/mol/\AA$^2$ was applied to the oxygen atoms of the QM waters to restrain their relative distances from the geometric center of the QM oxygens.
The initial $R_\mathrm{cut}$ was set to 16 \AA, and the subsequent $R_\mathrm{cut}$ was determined on the fly every 10 MD steps by searching starting from 13 \AA~with a step-size of 1 \AA~until two consecutive $R_\mathrm{cut}$ values both give an octupole error estimate under $\epsilon=2\times10^{-4}$ Hartree. $NVE$ dynamics was run for 99 ps after 1 ps of $NVT$ equilibration at 310 K. Figure~\ref{fig:econsv}B shows that the energy conservation of the QM/MM-Multipole dynamics is of a similar quality to that of classical force field dynamics, despite fluctuations in the number of MM atoms in Region I arising from the on-the-fly selection at every MD step of the regions using an $R_\mathrm{cut}$ that varies with time.

We further examine the energy conservation in the CM system as a more complicated example. $NVE$ dynamics was run for 10 ps after 250 fs $NVT$ equilibration under 300 K. The $R_\mathrm{cut}$ was dynamically determined every 5 MD steps by searching from 16 \AA~with a step-size of 2 \AA~until two consecutive $R_\mathrm{cut}$ values both give an error estimate under $\epsilon=2\times10^{-5}$ Hartree. The more frequent $R_\mathrm{cut}$ search and the tighter $\epsilon$ result in a larger fluctuation in the number of region I atoms. As shown in Figure~\ref{fig:econsv}C, the energy conservation of our QM/MM dynamics is again comparable to that of classical MD.

\subsection{Example Application: Catalytic Kinetics in Chorismate Mutase}
\begin{figure}
    \centering
    \includegraphics[width=15cm]{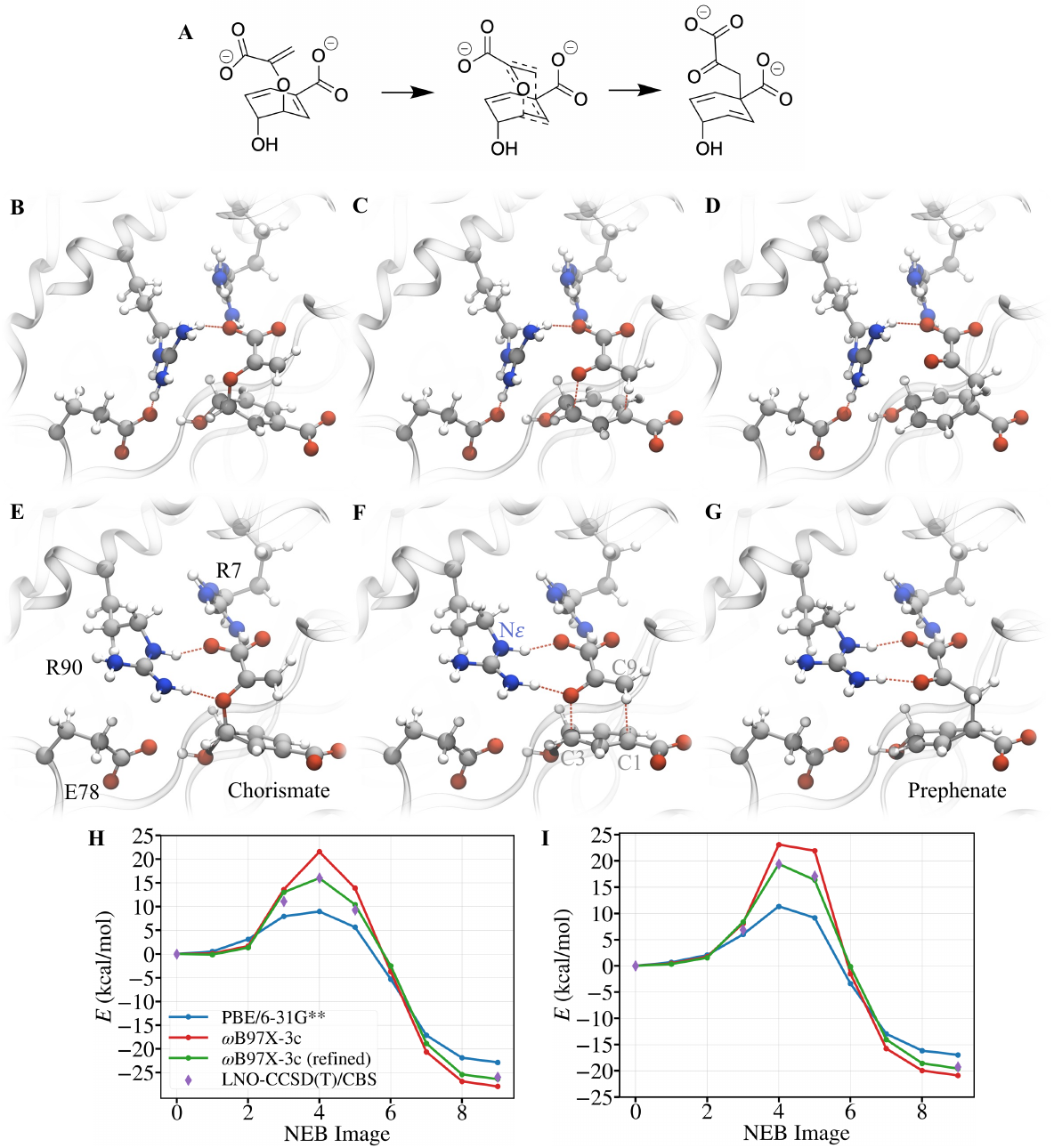}
    \caption{(A) The chorismate mutase catalyzed reaction. (B-D) Nudged elastic band optimized geometries of the reactant, the transition state, and the product when Arg90 forms a single hydrogen bond with the substrate. (E-G) The optimized reactant, transition state, and product geometries when Arg90 forms double hydrogen bonds with the substrate. \red{Energies are aligned at the reactant state.} (H) Gas-phase energies on NEB geometries for the 1HB conformation. (I) Gas-phase energies on NEB geometries for the 2HB conformation. }
    \label{fig:cm}
\end{figure}

\begin{table}[H]
    \centering
\resizebox{\textwidth}{!}{%
    \begin{tabular}{lllll}
    \hline
    Expt.  & \multicolumn{4}{c}{Theory} \\
     \cline{2-5}
    \multirow{4}{*}{16$\pm$14\textsuperscript{\emph{a}}} & Binding Mode/QM region &  PBE/6-31G**  &  $\omega$B97X-3c & $\omega$B97X-3c (refined)\\
&  1HB/S+R90         & $(5.1\pm0.3)\times10^3$        & $(2.7\pm0.4)\times10^{-8}$       & \red{$(4.0\pm0.7)\times10^{-3}$}   \\
&  2HB/S+R90         & \red{$(9.2\pm0.5)\times10^{5}$}&      $(1.1\pm0.2)\times10^{-4}$  &  \red{$(4\pm2)\times10^{-2}$}\\\
&  2HB/S+R90+R7+E78  & $(1.2\pm0.2)\times10^7$        &  $(5.8\pm0.7)\times10^{-4}$      & \red{$1.1\pm0.2$}\\
    \hline
    \multicolumn{5}{c}{ 
        \begin{minipage}{\textwidth}
         \textsuperscript{\emph{a}} The experimental catalytic rate constant at 300 K was computed from the fitted temperature-dependent rate constant $k_\mathrm{cat}=k_B T/h \exp{(-\Delta H/k_BT + \Delta S/k_B)}$ with parameters $\Delta H=12.7\pm0.4$ kcal/mol and $\Delta S=-9.1\pm1.2$ cal/mol/K, taken from Ref.\cite{kast1996chorismate}, divided by the solvent kinetic isotope effect (KIE) measured at 303.15 K\cite{guilford1987mechanism}. The error in this rate \red{constant} was estimated by the uncertainty propagation equation\cite{ku1966notes} by assuming the errors of $\Delta H$ and $\Delta S$ are independent and no error in KIE. The normalization by KIE is for direct comparison to the theoretical rate constants obtained with classical nuclear dynamics.
        \end{minipage}
    }
    \end{tabular}
}
    \caption{Catalytic rate constant of  chorismate mutase from \textit{Bacillus subtilis} in s$^{-1}$. Theoretical rate constant errors are reported as the square root of the rate constant estimator variance. Statistical analysis details are provided in the SI.}
    \label{tab:cm_rate}
\end{table}
Chorismate Mutase (CM) is a critical component in the primary and secondary metabolism of fungi, bacteria, and plants\cite{hubrich2021chorismate}. It catalyzes the Claisen rearrangement that transforms chorismate into prephenate (Figure~\ref{fig:cm}A), an important precursor for the biosynthesis of aromatic amino acids\cite{tzin2010new,noda2017recent,hubrich2021chorismate}. The catalytic kinetics of this reaction were recently studied by Ray et al. via MD enhanced sampling simulations on a QM(PBE)/MM PES\cite{ray}. They computed the reaction rate using a variant of the flooding method\cite{grubmuller1995predicting} called OPES flooding\cite{ray2022rare}, which adaptively adds bias potentials to the reactant state to accelerate barrier crossings that would otherwise be inaccessible to unbiased simulations. The computed QM/MM reaction rate constant was found to be somewhat higher than the experimentally measured catalytic rate constant, and this discrepancy was attributed by the authors to the neglect of the free energy cost for the substrate to adopt a ``reactive" conformation.

We now re-examine this reaction, following the simulation setup of Ray et al. as closely as possible, but further explore how the theoretical kinetic predictions depend on the MD sampling, the QM region definition, and the employed QM theory. In all our simulations, we used the same scheme to determine $R_\mathrm{cut}$ as in the energy conservation test. We verified our QM/MM scheme and implementation by reproducing the QM(PBE/DZVP)/MM result of CP2K by Ray et al. using our QM(PBE/6-31G**)/MM-Multipole MD.  We found that the calculated rate constant ($5.1\times10^3$ s$^{-1}$) indeed overestimates the experimental one of 16 s$^{-1}$ while being of the same order of magnitude as the value reported by Ray et al. (9433 s$^{-1}$).

We first consider how the rate constant depends on the level of quantum theory. To do so we used two more accurate hybrid density functionals: $\omega$B97X-3c and a refined version of it parametrized to reproduce the \red{LNO}-CCSD(T)/CBS reaction barriers for this system (see Computational Details). We kept the QM region definition unchanged and started from the same structure as Ray et al. (with appropriate equilibration using the corresponding functional). By changing the quantum description, we arrive at an opposite conclusion -- the QM/MM reaction rate constant now significantly underestimates the experimental catalytic rate constant (the first row of Table~\ref{tab:cm_rate}). 
The qualitatively different conclusion is a result of the inaccuracy in the pure functional treatment of the QM region, with its substantial underestimate of the reaction barrier, as shown by the gas-phase energetic analysis (Figure~\ref{fig:cm}H). Assuming that the estimated experimental rate constant is accurate and the refined $\omega$B97X-3c functional PES is almost exact, the deviation of the computed rate constant must be a result of either insufficient MD sampling (statistical error) or an inadequate QM region size (a source of systematic error).

As shown in Table~\ref{tab:cm_rate}, our calculated rate constant does not appear to suffer from a large statistical error
, but in the simulations initiated from the conformation of Ref.~\cite{ray}, we never observed a double hydrogen bond between the catalytic Arg90 and the substrate, even though this is a known binding mode found in a CM crystal structure (PDB id: 1COM)\cite{chook1994monofunctional}. 
Instead, Arg90 forms only one hydrogen bond with the substrate (Figure~\ref{fig:cm}B), resembling the conformation captured in another crystal structure (PDB id: 2CHT)\cite{chook1993crystal} \red{that was used to initiate the simulations of Ray et al}. The main difference between the two binding modes is in the orientations of the $\epsilon$-nitrogen/hydrogen of Arg90. In 2CHT, an HB between the Arg90 terminal nitrogen and the substrate carboxylic oxygen is found with no other HB interactions between the two moieties. In 1COM, two HBs are found, one between the $\epsilon$-nitrogen and the chorismate carboxylic oxygen, and another between the terminal nitrogen and the chorismate ester oxygen. The transition from the 1HB conformation to the 2HB conformation requires a flip of Arg90, which is not accessible in our QM/MM equilibration. 

To investigate how the Arg90 conformation affects the reaction rate, we added biasing potentials to artificially accelerate this flip and then equilibrated the resulting 2HB structure after removing the bias (see Figure~\ref{fig:cm}E for an example of the 2HB structures). 
We performed OPES flooding on this new structure to compute the rate constant. 
\red{We found then that the calculated rate constants were enhanced regardless of the functional choice, with the $\omega$B97X-3c one even increasing by four orders of magnitude (the second row of Table~\ref{tab:cm_rate}).}

The sensitivity of the rate to the local protein conformation underlines the challenges of converging the sampling, especially when one has no prior knowledge of the existence of alternative binding modes. 
Interestingly, the gas-phase single point energetics show a lower reaction barrier for the 1HB binding mode than for the 2HB binding mode (Figures~\ref{fig:cm}H and~\ref{fig:cm}I), contradicting 
\red{the faster reaction rate constants} obtained from OPES flooding simulations with the 2HB conformation.
This illustrates the importance of entropic contributions and the appropriate modeling of the chemical environment.

It has been conjectured that the catalytic mechanism of CM involves an acid/base attack step due to the considerable measured solvent kinetic isotope effects (KIE)\cite{guilford1987mechanism} on the apparent catalytic rate constant, $k_\mathrm{H_2O}/k_\mathrm{D_2O}=2.23$. The 
\red{sensitivity} of the theoretical rate constant on the hydrogen-bonding pattern suggests, however, that the solvent KIE results from the hydrogen bond strength's dependence on the quantum nature of the protons; further simulations with an explicit treatment of nuclear quantum effects (NQEs) should be carried out to explore this point.

We also tested the sensitivity of the rate constant to the definition of the QM region by further including Arg7 and Glu78 in the QM region. When switching these residues from an MM to a QM description, 
\red{we found all the calculated rate constants  further increase (the third row of Table~\ref{tab:cm_rate}), consistent with substantially lower enzyme activities in mutagenesis analyses of Arg7 and Glu78 \cite{kast1996electrostatic,liu1996analysis}. The QM treatment allows a proton on Arg90 to be shared with Glu78 in PBE/6-31G** simulations, and even occasionally, for a proton transfer (PT) from Arg90 to Glu78 (1 in 11 simulations of $\omega$B97X-3c and 4 in 11 simulations of the refined $\omega$B97X-3c). However, the difference in the $\ln{k}$ of the refined $\omega$B97X-3c with 2HB/S+R90+R7+E78 is not statistically significant between the runs where PT happens and the ones where it does not (the $p$-value from a Welch's $t$-test is 0.36 assuming equal means as the null hypothesis). This lack of statistical significance suggests that the enhanced rate cannot be directly attributed to the occurrence of PT alone. Instead, the observed rate increase may be explained by other factors, including the potential charge transfer and the differences in the electrostatic representation between QM and classical MM charges.}

Our final theoretical rate constant with the refined $\omega$B97X-3c functional, the (larger) S+R90+R7+E78 QM region definition, and the double hydrogen-bonded Arg90 conformation, shows a reasonable agreement with the estimated experimental one. At the same time, our simulations illustrate the difficulties of obtaining a fully converged result. 
Improving this agreement will require improving the errors in the theoretical treatment, from the imperfect electronic structure, size of QM region, sampling, nuclear quantum effects, \red{and probably the rate theory as well (the correct estimate of rates depends on a careful choice of OPES flooding parameters according to Ref.~\cite{ray2022rare})}.
Within DFT treatments, the quality of the functional remains perhaps the largest source of uncertainty, with almost 11 orders of magnitude difference in rate constant between different functionals.


From the PBE data, where the rate constant with the 2HB Arg90 conformation and the larger QM region is much faster than the experimental value, one might conclude that the Claisen rearrangement step is not rate-limiting for the whole catalytic cycle, and perhaps the substrate leaving step is instead rate-limiting, as has been hypothesized in some earlier experiments\cite{kast1996chorismate}. However, 
our results using the improved PES provided by the refined $\omega$B97X-3C functional suggest the chemical transformation step is rate-limiting regardless of the Arg90 conformation or the QM region definition. This aligns with the proposal that electrostatic transition state stabilization is fundamental for CM catalysis\cite{burschowsky2014electrostatic}, and consistent with data that finds impaired enzyme activity after mutating Arg90 into a neutral citrulline, an arginine analog that is still able to preorganize the chorismate into its ``reactive" conformation.

\section{Conclusions}
In this work, we described a QM/MM implementation with an improved treatment of the periodic QM/MM electrostatics which leverages recent advances in GPU DFT algorithms to enable the practical use of advanced hybrid density functionals. Our benchmarks verify its numerical stability and illustrate the possibility of controlling and converging the error associated with the electrostatics. 

While this work has not focused on achieving a computationally optimal implementation, the open-source availability of our code and the already competitive level of performance achieved form a foundation for broader community efforts to build an even faster QM/MM code.


Our QM/MM investigation of the chorismate mutase reaction highlights the critical importance of converged conformational sampling and an accurate QM treatment in theoretical kinetic studies of enzymatic reactions. Despite the simplicity of this system, it appears challenging to obtain a fully converged theoretical description.  
We view this work as a first step towards improving the quality of sampling and QM modeling as part of our ongoing efforts to obtain a precise understanding of enzymatic reactions from direct quantum mechanical simulations.


\begin{suppinfo}
Supporting Information available. Details on multipolar Ewald and pseudo-bond parameterization, additional computational details on chorismate mutase, rate calculations, and statistical error analysis.
\end{suppinfo}

\section{Author Contributions} CL and GKC formulated the project. CL carried out the implementation and calculations. CL and GKC discussed the results and interpretation and wrote the manuscript.

\begin{acknowledgement}
This work was primarily supported by the US Department of Energy, Office of Science, Basic Energy Sciences, through  Award No. DE-SC0023318. GKC acknowledges additional support in the conceptualization phase from the Dreyfus Foundation, under the program Machine Learning in the Chemical Sciences and Engineering, and from the Simons Investigator program.
\end{acknowledgement}

\bibliography{ref}

\providecommand{\latin}[1]{#1}
\makeatletter
\providecommand{\doi}
  {\begingroup\let\do\@makeother\dospecials
  \catcode`\{=1 \catcode`\}=2 \doi@aux}
\providecommand{\doi@aux}[1]{\endgroup\texttt{#1}}
\makeatother
\providecommand*\mcitethebibliography{\thebibliography}
\csname @ifundefined\endcsname{endmcitethebibliography}  {\let\endmcitethebibliography\endthebibliography}{}
\begin{mcitethebibliography}{104}
\providecommand*\natexlab[1]{#1}
\providecommand*\mciteSetBstSublistMode[1]{}
\providecommand*\mciteSetBstMaxWidthForm[2]{}
\providecommand*\mciteBstWouldAddEndPuncttrue
  {\def\EndOfBibitem{\unskip.}}
\providecommand*\mciteBstWouldAddEndPunctfalse
  {\let\EndOfBibitem\relax}
\providecommand*\mciteSetBstMidEndSepPunct[3]{}
\providecommand*\mciteSetBstSublistLabelBeginEnd[3]{}
\providecommand*\EndOfBibitem{}
\mciteSetBstSublistMode{f}
\mciteSetBstMaxWidthForm{subitem}{(\alph{mcitesubitemcount})}
\mciteSetBstSublistLabelBeginEnd
  {\mcitemaxwidthsubitemform\space}
  {\relax}
  {\relax}

\bibitem[Warshel and Levitt(1976)Warshel, and Levitt]{warshel1976theoretical}
Warshel,~A.; Levitt,~M. Theoretical studies of enzymic reactions: dielectric, electrostatic and steric stabilization of the carbonium ion in the reaction of lysozyme. \emph{Journal of molecular biology} \textbf{1976}, \emph{103}, 227--249\relax
\mciteBstWouldAddEndPuncttrue
\mciteSetBstMidEndSepPunct{\mcitedefaultmidpunct}
{\mcitedefaultendpunct}{\mcitedefaultseppunct}\relax
\EndOfBibitem
\bibitem[Field \latin{et~al.}(1990)Field, Bash, and Karplus]{karplus1990combined}
Field,~M.~J.; Bash,~P.~A.; Karplus,~M. A combined quantum mechanical and molecular mechanical potential for molecular dynamics simulations. \emph{Journal of computational chemistry} \textbf{1990}, \emph{11}, 700--733\relax
\mciteBstWouldAddEndPuncttrue
\mciteSetBstMidEndSepPunct{\mcitedefaultmidpunct}
{\mcitedefaultendpunct}{\mcitedefaultseppunct}\relax
\EndOfBibitem
\bibitem[Gao(1996)]{gao1996methods}
Gao,~J. Methods and applications of combined quantum mechanical and molecular mechanical potentials. \emph{Reviews in computational chemistry} \textbf{1996}, 119--185\relax
\mciteBstWouldAddEndPuncttrue
\mciteSetBstMidEndSepPunct{\mcitedefaultmidpunct}
{\mcitedefaultendpunct}{\mcitedefaultseppunct}\relax
\EndOfBibitem
\bibitem[Hu and Yang(2008)Hu, and Yang]{hu2008free}
Hu,~H.; Yang,~W. Free energies of chemical reactions in solution and in enzymes with ab initio quantum mechanics/molecular mechanics methods. \emph{Annu. Rev. Phys. Chem.} \textbf{2008}, \emph{59}, 573--601\relax
\mciteBstWouldAddEndPuncttrue
\mciteSetBstMidEndSepPunct{\mcitedefaultmidpunct}
{\mcitedefaultendpunct}{\mcitedefaultseppunct}\relax
\EndOfBibitem
\bibitem[Chung \latin{et~al.}(2015)Chung, Sameera, Ramozzi, Page, Hatanaka, Petrova, Harris, Li, Ke, Liu, \latin{et~al.} others]{chung2015oniom}
Chung,~L.~W.; Sameera,~W.; Ramozzi,~R.; Page,~A.~J.; Hatanaka,~M.; Petrova,~G.~P.; Harris,~T.~V.; Li,~X.; Ke,~Z.; Liu,~F.; others The ONIOM method and its applications. \emph{Chemical reviews} \textbf{2015}, \emph{115}, 5678--5796\relax
\mciteBstWouldAddEndPuncttrue
\mciteSetBstMidEndSepPunct{\mcitedefaultmidpunct}
{\mcitedefaultendpunct}{\mcitedefaultseppunct}\relax
\EndOfBibitem
\bibitem[Senn and Thiel(2009)Senn, and Thiel]{senn2009qm}
Senn,~H.~M.; Thiel,~W. QM/MM methods for biomolecular systems. \emph{Angewandte Chemie International Edition} \textbf{2009}, \emph{48}, 1198--1229\relax
\mciteBstWouldAddEndPuncttrue
\mciteSetBstMidEndSepPunct{\mcitedefaultmidpunct}
{\mcitedefaultendpunct}{\mcitedefaultseppunct}\relax
\EndOfBibitem
\bibitem[Brunk and Rothlisberger(2015)Brunk, and Rothlisberger]{brunk2015mixed}
Brunk,~E.; Rothlisberger,~U. Mixed quantum mechanical/molecular mechanical molecular dynamics simulations of biological systems in ground and electronically excited states. \emph{Chemical reviews} \textbf{2015}, \emph{115}, 6217--6263\relax
\mciteBstWouldAddEndPuncttrue
\mciteSetBstMidEndSepPunct{\mcitedefaultmidpunct}
{\mcitedefaultendpunct}{\mcitedefaultseppunct}\relax
\EndOfBibitem
\bibitem[Lonsdale and J~Mulholland(2014)Lonsdale, and J~Mulholland]{lonsdale2014qm}
Lonsdale,~R.; J~Mulholland,~A. QM/MM modelling of drug-metabolizing enzymes. \emph{Current Topics in Medicinal Chemistry} \textbf{2014}, \emph{14}, 1339--1347\relax
\mciteBstWouldAddEndPuncttrue
\mciteSetBstMidEndSepPunct{\mcitedefaultmidpunct}
{\mcitedefaultendpunct}{\mcitedefaultseppunct}\relax
\EndOfBibitem
\bibitem[Cui \latin{et~al.}(2021)Cui, Pal, and Xie]{cui2021biomolecular}
Cui,~Q.; Pal,~T.; Xie,~L. Biomolecular QM/MM simulations: What are some of the “burning issues”? \emph{The Journal of Physical Chemistry B} \textbf{2021}, \emph{125}, 689--702\relax
\mciteBstWouldAddEndPuncttrue
\mciteSetBstMidEndSepPunct{\mcitedefaultmidpunct}
{\mcitedefaultendpunct}{\mcitedefaultseppunct}\relax
\EndOfBibitem
\bibitem[Kuba{\v{r}} \latin{et~al.}(2023)Kuba{\v{r}}, Elstner, and Cui]{kubavr2023hybrid}
Kuba{\v{r}},~T.; Elstner,~M.; Cui,~Q. Hybrid quantum mechanical/molecular mechanical methods for studying energy transduction in biomolecular machines. \emph{Annual Review of Biophysics} \textbf{2023}, \emph{52}, 525--551\relax
\mciteBstWouldAddEndPuncttrue
\mciteSetBstMidEndSepPunct{\mcitedefaultmidpunct}
{\mcitedefaultendpunct}{\mcitedefaultseppunct}\relax
\EndOfBibitem
\bibitem[Li \latin{et~al.}(2024)Li, Sun, Zhang, and Chan]{li2024introducting}
Li,~R.; Sun,~Q.; Zhang,~X.; Chan,~G. K.-L. Introducing GPU-acceleration into the Python-based Simulations of Chemistry Framework. 2024; \url{https://arxiv.org/abs/2407.09700}\relax
\mciteBstWouldAddEndPuncttrue
\mciteSetBstMidEndSepPunct{\mcitedefaultmidpunct}
{\mcitedefaultendpunct}{\mcitedefaultseppunct}\relax
\EndOfBibitem
\bibitem[Wu \latin{et~al.}(2024)Wu, Sun, Pu, Zheng, Ma, Yan, Yu, Wu, Huo, Li, Ren, Gong, Zhang, and Gao]{wu2024enhancing}
Wu,~X.; Sun,~Q.; Pu,~Z.; Zheng,~T.; Ma,~W.; Yan,~W.; Yu,~X.; Wu,~Z.; Huo,~M.; Li,~X.; Ren,~W.; Gong,~S.; Zhang,~Y.; Gao,~W. Enhancing GPU-acceleration in the Python-based Simulations of Chemistry Framework. 2024; \url{https://arxiv.org/abs/2404.09452}\relax
\mciteBstWouldAddEndPuncttrue
\mciteSetBstMidEndSepPunct{\mcitedefaultmidpunct}
{\mcitedefaultendpunct}{\mcitedefaultseppunct}\relax
\EndOfBibitem
\bibitem[R{\'a}k and Cserey(2015)R{\'a}k, and Cserey]{rak2015brush}
R{\'a}k,~{\'A}.; Cserey,~G. The BRUSH algorithm for two-electron integrals on GPU. \emph{Chemical Physics Letters} \textbf{2015}, \emph{622}, 92--98\relax
\mciteBstWouldAddEndPuncttrue
\mciteSetBstMidEndSepPunct{\mcitedefaultmidpunct}
{\mcitedefaultendpunct}{\mcitedefaultseppunct}\relax
\EndOfBibitem
\bibitem[Brooks~III \latin{et~al.}(1985)Brooks~III, Pettitt, and Karplus]{brooks1985structural}
Brooks~III,~C.~L.; Pettitt,~B.~M.; Karplus,~M. Structural and energetic effects of truncating long ranged interactions in ionic and polar fluids. \emph{The Journal of chemical physics} \textbf{1985}, \emph{83}, 5897--5908\relax
\mciteBstWouldAddEndPuncttrue
\mciteSetBstMidEndSepPunct{\mcitedefaultmidpunct}
{\mcitedefaultendpunct}{\mcitedefaultseppunct}\relax
\EndOfBibitem
\bibitem[York \latin{et~al.}(1995)York, Yang, Lee, Darden, and Pedersen]{york1995toward}
York,~D.~M.; Yang,~W.; Lee,~H.; Darden,~T.; Pedersen,~L.~G. Toward the accurate modeling of DNA: the importance of long-range electrostatics. \emph{Journal of the American Chemical Society} \textbf{1995}, \emph{117}, 5001--5002\relax
\mciteBstWouldAddEndPuncttrue
\mciteSetBstMidEndSepPunct{\mcitedefaultmidpunct}
{\mcitedefaultendpunct}{\mcitedefaultseppunct}\relax
\EndOfBibitem
\bibitem[Sanz-Navarro \latin{et~al.}(2011)Sanz-Navarro, Grima, Garc{\'\i}a, Bea, Soba, Cela, and Ordej{\'o}n]{sanz2011efficient}
Sanz-Navarro,~C.~F.; Grima,~R.; Garc{\'\i}a,~A.; Bea,~E.~A.; Soba,~A.; Cela,~J.~M.; Ordej{\'o}n,~P. An efficient implementation of a QM--MM method in SIESTA. \emph{Theoretical Chemistry Accounts} \textbf{2011}, \emph{128}, 825--833\relax
\mciteBstWouldAddEndPuncttrue
\mciteSetBstMidEndSepPunct{\mcitedefaultmidpunct}
{\mcitedefaultendpunct}{\mcitedefaultseppunct}\relax
\EndOfBibitem
\bibitem[Kawashima \latin{et~al.}(2019)Kawashima, Ishimura, and Shiga]{kawashima2019ab}
Kawashima,~Y.; Ishimura,~K.; Shiga,~M. Ab initio quantum mechanics/molecular mechanics method with periodic boundaries employing Ewald summation technique to electron-charge interaction: Treatment of the surface-dipole term. \emph{The Journal of Chemical Physics} \textbf{2019}, \emph{150}, 124103\relax
\mciteBstWouldAddEndPuncttrue
\mciteSetBstMidEndSepPunct{\mcitedefaultmidpunct}
{\mcitedefaultendpunct}{\mcitedefaultseppunct}\relax
\EndOfBibitem
\bibitem[VandeVondele \latin{et~al.}(2005)VandeVondele, Krack, Mohamed, Parrinello, Chassaing, and Hutter]{vandevondele2005quickstep}
VandeVondele,~J.; Krack,~M.; Mohamed,~F.; Parrinello,~M.; Chassaing,~T.; Hutter,~J. Quickstep: Fast and accurate density functional calculations using a mixed Gaussian and plane waves approach. \emph{Computer Physics Communications} \textbf{2005}, \emph{167}, 103--128\relax
\mciteBstWouldAddEndPuncttrue
\mciteSetBstMidEndSepPunct{\mcitedefaultmidpunct}
{\mcitedefaultendpunct}{\mcitedefaultseppunct}\relax
\EndOfBibitem
\bibitem[Laino \latin{et~al.}(2005)Laino, Mohamed, Laio, and Parrinello]{laino2005efficient}
Laino,~T.; Mohamed,~F.; Laio,~A.; Parrinello,~M. An efficient real space multigrid QM/MM electrostatic coupling. \emph{Journal of Chemical Theory and Computation} \textbf{2005}, \emph{1}, 1176--1184\relax
\mciteBstWouldAddEndPuncttrue
\mciteSetBstMidEndSepPunct{\mcitedefaultmidpunct}
{\mcitedefaultendpunct}{\mcitedefaultseppunct}\relax
\EndOfBibitem
\bibitem[Laino \latin{et~al.}(2006)Laino, Mohamed, Laio, and Parrinello]{laino2006efficient}
Laino,~T.; Mohamed,~F.; Laio,~A.; Parrinello,~M. An efficient linear-scaling electrostatic coupling for treating periodic boundary conditions in QM/MM simulations. \emph{Journal of chemical theory and computation} \textbf{2006}, \emph{2}, 1370--1378\relax
\mciteBstWouldAddEndPuncttrue
\mciteSetBstMidEndSepPunct{\mcitedefaultmidpunct}
{\mcitedefaultendpunct}{\mcitedefaultseppunct}\relax
\EndOfBibitem
\bibitem[Giese and York(2016)Giese, and York]{giese2016ambient}
Giese,~T.~J.; York,~D.~M. Ambient-potential composite Ewald method for ab initio quantum mechanical/molecular mechanical molecular dynamics simulation. \emph{Journal of chemical theory and computation} \textbf{2016}, \emph{12}, 2611--2632\relax
\mciteBstWouldAddEndPuncttrue
\mciteSetBstMidEndSepPunct{\mcitedefaultmidpunct}
{\mcitedefaultendpunct}{\mcitedefaultseppunct}\relax
\EndOfBibitem
\bibitem[Pederson and McDaniel(2022)Pederson, and McDaniel]{pederson2022dft}
Pederson,~J.~P.; McDaniel,~J.~G. DFT-based QM/MM with particle-mesh Ewald for direct, long-range electrostatic embedding. \emph{The Journal of Chemical Physics} \textbf{2022}, \emph{156}, 174105\relax
\mciteBstWouldAddEndPuncttrue
\mciteSetBstMidEndSepPunct{\mcitedefaultmidpunct}
{\mcitedefaultendpunct}{\mcitedefaultseppunct}\relax
\EndOfBibitem
\bibitem[Nam \latin{et~al.}(2005)Nam, Gao, and York]{nam2005efficient}
Nam,~K.; Gao,~J.; York,~D.~M. An efficient linear-scaling Ewald method for long-range electrostatic interactions in combined QM/MM calculations. \emph{Journal of Chemical Theory and Computation} \textbf{2005}, \emph{1}, 2--13\relax
\mciteBstWouldAddEndPuncttrue
\mciteSetBstMidEndSepPunct{\mcitedefaultmidpunct}
{\mcitedefaultendpunct}{\mcitedefaultseppunct}\relax
\EndOfBibitem
\bibitem[Riccardi \latin{et~al.}(2005)Riccardi, Schaefer, and Cui]{riccardi2005p}
Riccardi,~D.; Schaefer,~P.; Cui,~Q. p K a calculations in solution and proteins with QM/MM free energy perturbation simulations: A quantitative test of QM/MM protocols. \emph{The Journal of Physical Chemistry B} \textbf{2005}, \emph{109}, 17715--17733\relax
\mciteBstWouldAddEndPuncttrue
\mciteSetBstMidEndSepPunct{\mcitedefaultmidpunct}
{\mcitedefaultendpunct}{\mcitedefaultseppunct}\relax
\EndOfBibitem
\bibitem[Seabra \latin{et~al.}(2007)Seabra, Walker, Elstner, Case, and Roitberg]{seabra2007implementation}
Seabra,~G. d.~M.; Walker,~R.~C.; Elstner,~M.; Case,~D.~A.; Roitberg,~A.~E. Implementation of the SCC-DFTB method for hybrid QM/MM simulations within the Amber molecular dynamics package. \emph{The Journal of Physical Chemistry A} \textbf{2007}, \emph{111}, 5655--5664\relax
\mciteBstWouldAddEndPuncttrue
\mciteSetBstMidEndSepPunct{\mcitedefaultmidpunct}
{\mcitedefaultendpunct}{\mcitedefaultseppunct}\relax
\EndOfBibitem
\bibitem[Walker \latin{et~al.}(2008)Walker, Crowley, and Case]{walker2008implementation}
Walker,~R.~C.; Crowley,~M.~F.; Case,~D.~A. The implementation of a fast and accurate QM/MM potential method in Amber. \emph{Journal of computational chemistry} \textbf{2008}, \emph{29}, 1019--1031\relax
\mciteBstWouldAddEndPuncttrue
\mciteSetBstMidEndSepPunct{\mcitedefaultmidpunct}
{\mcitedefaultendpunct}{\mcitedefaultseppunct}\relax
\EndOfBibitem
\bibitem[Holden \latin{et~al.}(2013)Holden, Richard, and Herbert]{holden2013periodic}
Holden,~Z.~C.; Richard,~R.~M.; Herbert,~J.~M. Periodic boundary conditions for QM/MM calculations: Ewald summation for extended Gaussian basis sets. \emph{The Journal of Chemical Physics} \textbf{2013}, \emph{139}, 244108\relax
\mciteBstWouldAddEndPuncttrue
\mciteSetBstMidEndSepPunct{\mcitedefaultmidpunct}
{\mcitedefaultendpunct}{\mcitedefaultseppunct}\relax
\EndOfBibitem
\bibitem[Kuba{\v{r}} \latin{et~al.}(2015)Kuba{\v{r}}, Welke, and Groenhof]{kubavr2015new}
Kuba{\v{r}},~T.; Welke,~K.; Groenhof,~G. New QM/MM implementation of the DFTB3 method in the gromacs package. 2015\relax
\mciteBstWouldAddEndPuncttrue
\mciteSetBstMidEndSepPunct{\mcitedefaultmidpunct}
{\mcitedefaultendpunct}{\mcitedefaultseppunct}\relax
\EndOfBibitem
\bibitem[Nishizawa and Okumura(2016)Nishizawa, and Okumura]{nishizawa2016rapid}
Nishizawa,~H.; Okumura,~H. Rapid QM/MM approach for biomolecular systems under periodic boundary conditions: Combination of the density-functional tight-binding theory and particle mesh Ewald method. \emph{Journal of Computational Chemistry} \textbf{2016}, \emph{37}, 2701--2711\relax
\mciteBstWouldAddEndPuncttrue
\mciteSetBstMidEndSepPunct{\mcitedefaultmidpunct}
{\mcitedefaultendpunct}{\mcitedefaultseppunct}\relax
\EndOfBibitem
\bibitem[Holden \latin{et~al.}(2019)Holden, Rana, and Herbert]{holden2019analytic}
Holden,~Z.~C.; Rana,~B.; Herbert,~J.~M. Analytic gradient for the QM/MM-Ewald method using charges derived from the electrostatic potential: Theory, implementation, and application to ab initio molecular dynamics simulation of the aqueous electron. \emph{The Journal of Chemical Physics} \textbf{2019}, \emph{150}, 144115\relax
\mciteBstWouldAddEndPuncttrue
\mciteSetBstMidEndSepPunct{\mcitedefaultmidpunct}
{\mcitedefaultendpunct}{\mcitedefaultseppunct}\relax
\EndOfBibitem
\bibitem[Bonfrate \latin{et~al.}(2023)Bonfrate, Ferr{\'e}, and Huix-Rotllant]{bonfrate2023efficient}
Bonfrate,~S.; Ferr{\'e},~N.; Huix-Rotllant,~M. An efficient electrostatic embedding QM/MM method using periodic boundary conditions based on particle-mesh Ewald sums and electrostatic potential fitted charge operators. \emph{The Journal of Chemical Physics} \textbf{2023}, \emph{158}\relax
\mciteBstWouldAddEndPuncttrue
\mciteSetBstMidEndSepPunct{\mcitedefaultmidpunct}
{\mcitedefaultendpunct}{\mcitedefaultseppunct}\relax
\EndOfBibitem
\bibitem[Holden \latin{et~al.}(2015)Holden, Richard, and Herbert]{holden2015erratum}
Holden,~Z.~C.; Richard,~R.~M.; Herbert,~J.~M. Erratum:“Periodic boundary conditions for QM/MM calculations: Ewald summation for extended Gaussian basis sets”[J. Chem. Phys. 139, 244108 (2013)]. \emph{The Journal of Chemical Physics} \textbf{2015}, \emph{142}, 059901\relax
\mciteBstWouldAddEndPuncttrue
\mciteSetBstMidEndSepPunct{\mcitedefaultmidpunct}
{\mcitedefaultendpunct}{\mcitedefaultseppunct}\relax
\EndOfBibitem
\bibitem[Laio \latin{et~al.}(2002)Laio, VandeVondele, and Rothlisberger]{laio2002hamiltonian}
Laio,~A.; VandeVondele,~J.; Rothlisberger,~U. A Hamiltonian electrostatic coupling scheme for hybrid Car--Parrinello molecular dynamics simulations. \emph{The Journal of chemical physics} \textbf{2002}, \emph{116}, 6941--6947\relax
\mciteBstWouldAddEndPuncttrue
\mciteSetBstMidEndSepPunct{\mcitedefaultmidpunct}
{\mcitedefaultendpunct}{\mcitedefaultseppunct}\relax
\EndOfBibitem
\bibitem[Janowski \latin{et~al.}(2012)Janowski, Wolinski, and Pulay]{janowski2012ultrafast}
Janowski,~T.; Wolinski,~K.; Pulay,~P. Ultrafast quantum mechanics/molecular mechanics Monte Carlo simulations using generalized multipole polarizabilities. \emph{Chemical Physics Letters} \textbf{2012}, \emph{530}, 1--9\relax
\mciteBstWouldAddEndPuncttrue
\mciteSetBstMidEndSepPunct{\mcitedefaultmidpunct}
{\mcitedefaultendpunct}{\mcitedefaultseppunct}\relax
\EndOfBibitem
\bibitem[Alvarez-Ibarra \latin{et~al.}(2012)Alvarez-Ibarra, Köster, Zhang, and Salahub]{alvarez2012asymptotic}
Alvarez-Ibarra,~A.; Köster,~A.~M.; Zhang,~R.; Salahub,~D.~R. Asymptotic expansion for electrostatic embedding integrals in QM/MM calculations. \emph{Journal of Chemical Theory and Computation} \textbf{2012}, \emph{8}, 4232--4238\relax
\mciteBstWouldAddEndPuncttrue
\mciteSetBstMidEndSepPunct{\mcitedefaultmidpunct}
{\mcitedefaultendpunct}{\mcitedefaultseppunct}\relax
\EndOfBibitem
\bibitem[Giese \latin{et~al.}(2015)Giese, Panteva, Chen, and York]{giese2015multipolar}
Giese,~T.~J.; Panteva,~M.~T.; Chen,~H.; York,~D.~M. Multipolar Ewald methods, 1: Theory, accuracy, and performance. \emph{Journal of Chemical Theory and Computation} \textbf{2015}, \emph{11}, 436--450\relax
\mciteBstWouldAddEndPuncttrue
\mciteSetBstMidEndSepPunct{\mcitedefaultmidpunct}
{\mcitedefaultendpunct}{\mcitedefaultseppunct}\relax
\EndOfBibitem
\bibitem[Dziedzic \latin{et~al.}(2016)Dziedzic, Mao, Shao, Ponder, Head-Gordon, Head-Gordon, and Skylaris]{dziedzic2016tinktep}
Dziedzic,~J.; Mao,~Y.; Shao,~Y.; Ponder,~J.; Head-Gordon,~T.; Head-Gordon,~M.; Skylaris,~C.-K. TINKTEP: A fully self-consistent, mutually polarizable QM/MM approach based on the AMOEBA force field. \emph{The Journal of chemical physics} \textbf{2016}, \emph{145}, 124106\relax
\mciteBstWouldAddEndPuncttrue
\mciteSetBstMidEndSepPunct{\mcitedefaultmidpunct}
{\mcitedefaultendpunct}{\mcitedefaultseppunct}\relax
\EndOfBibitem
\bibitem[Pan \latin{et~al.}(2018)Pan, Rosta, and Shao]{pan2018representation}
Pan,~X.; Rosta,~E.; Shao,~Y. Representation of the QM subsystem for long-range electrostatic interaction in non-periodic ab initio QM/MM calculations. \emph{Molecules} \textbf{2018}, \emph{23}, 2500\relax
\mciteBstWouldAddEndPuncttrue
\mciteSetBstMidEndSepPunct{\mcitedefaultmidpunct}
{\mcitedefaultendpunct}{\mcitedefaultseppunct}\relax
\EndOfBibitem
\bibitem[Olsen \latin{et~al.}(2019)Olsen, Bolnykh, Meloni, Ippoliti, Bircher, Carloni, and Rothlisberger]{olsen2019mimic}
Olsen,~J. M.~H.; Bolnykh,~V.; Meloni,~S.; Ippoliti,~E.; Bircher,~M.~P.; Carloni,~P.; Rothlisberger,~U. MiMiC: a novel framework for multiscale modeling in computational chemistry. \emph{Journal of chemical theory and computation} \textbf{2019}, \emph{15}, 3810--3823\relax
\mciteBstWouldAddEndPuncttrue
\mciteSetBstMidEndSepPunct{\mcitedefaultmidpunct}
{\mcitedefaultendpunct}{\mcitedefaultseppunct}\relax
\EndOfBibitem
\bibitem[Bolnykh \latin{et~al.}(2019)Bolnykh, Olsen, Meloni, Bircher, Ippoliti, Carloni, and Rothlisberger]{bolnykh2019extreme}
Bolnykh,~V.; Olsen,~J. M.~H.; Meloni,~S.; Bircher,~M.~P.; Ippoliti,~E.; Carloni,~P.; Rothlisberger,~U. Extreme scalability of DFT-based QM/MM MD simulations using MiMiC. \emph{Journal of chemical theory and computation} \textbf{2019}, \emph{15}, 5601--5613\relax
\mciteBstWouldAddEndPuncttrue
\mciteSetBstMidEndSepPunct{\mcitedefaultmidpunct}
{\mcitedefaultendpunct}{\mcitedefaultseppunct}\relax
\EndOfBibitem
\bibitem[Dziedzic \latin{et~al.}(2019)Dziedzic, Head-Gordon, Head-Gordon, and Skylaris]{dziedzic2019mutually}
Dziedzic,~J.; Head-Gordon,~T.; Head-Gordon,~M.; Skylaris,~C.-K. Mutually polarizable QM/MM model with in situ optimized localized basis functions. \emph{The Journal of Chemical Physics} \textbf{2019}, \emph{150}, 074103\relax
\mciteBstWouldAddEndPuncttrue
\mciteSetBstMidEndSepPunct{\mcitedefaultmidpunct}
{\mcitedefaultendpunct}{\mcitedefaultseppunct}\relax
\EndOfBibitem
\bibitem[Kirsch \latin{et~al.}(2021)Kirsch, Olsen, Bolnykh, Meloni, Ippoliti, Rothlisberger, Cascella, and Gauss]{kirsch2021wavefunction}
Kirsch,~T.; Olsen,~J. M.~H.; Bolnykh,~V.; Meloni,~S.; Ippoliti,~E.; Rothlisberger,~U.; Cascella,~M.; Gauss,~J. Wavefunction-based electrostatic-embedding QM/MM using CFOUR through MiMiC. \emph{Journal of Chemical Theory and Computation} \textbf{2021}, \emph{18}, 13--24\relax
\mciteBstWouldAddEndPuncttrue
\mciteSetBstMidEndSepPunct{\mcitedefaultmidpunct}
{\mcitedefaultendpunct}{\mcitedefaultseppunct}\relax
\EndOfBibitem
\bibitem[Pan \latin{et~al.}(2021)Pan, Nam, Epifanovsky, Simmonett, Rosta, and Shao]{pan2021simplified}
Pan,~X.; Nam,~K.; Epifanovsky,~E.; Simmonett,~A.~C.; Rosta,~E.; Shao,~Y. A simplified charge projection scheme for long-range electrostatics in ab initio QM/MM calculations. \emph{The Journal of Chemical Physics} \textbf{2021}, \emph{154}, 024115\relax
\mciteBstWouldAddEndPuncttrue
\mciteSetBstMidEndSepPunct{\mcitedefaultmidpunct}
{\mcitedefaultendpunct}{\mcitedefaultseppunct}\relax
\EndOfBibitem
\bibitem[Reinholdt \latin{et~al.}(2021)Reinholdt, Kongsted, and Lipparini]{reinholdt2021fast}
Reinholdt,~P.; Kongsted,~J.; Lipparini,~F. Fast approximate but accurate QM/MM interactions for polarizable embedding. \emph{Journal of Chemical theory and Computation} \textbf{2021}, \emph{18}, 344--356\relax
\mciteBstWouldAddEndPuncttrue
\mciteSetBstMidEndSepPunct{\mcitedefaultmidpunct}
{\mcitedefaultendpunct}{\mcitedefaultseppunct}\relax
\EndOfBibitem
\bibitem[Polonius \latin{et~al.}(2023)Polonius, Zhuravel, Bachmair, and Mai]{polonius2023lvc}
Polonius,~S.; Zhuravel,~O.; Bachmair,~B.; Mai,~S. LVC/MM: A Hybrid Linear Vibronic Coupling/Molecular Mechanics Model with Distributed Multipole-Based Electrostatic Embedding for Highly Efficient Surface Hopping Dynamics in Solution. \emph{Journal of Chemical Theory and Computation} \textbf{2023}, \emph{19}, 7171--7186\relax
\mciteBstWouldAddEndPuncttrue
\mciteSetBstMidEndSepPunct{\mcitedefaultmidpunct}
{\mcitedefaultendpunct}{\mcitedefaultseppunct}\relax
\EndOfBibitem
\bibitem[Frenkel and Smit(2023)Frenkel, and Smit]{frenkel2023understanding}
Frenkel,~D.; Smit,~B. \emph{Understanding molecular simulation: from algorithms to applications}; Elsevier, 2023\relax
\mciteBstWouldAddEndPuncttrue
\mciteSetBstMidEndSepPunct{\mcitedefaultmidpunct}
{\mcitedefaultendpunct}{\mcitedefaultseppunct}\relax
\EndOfBibitem
\bibitem[Sokalski and Poirier(1983)Sokalski, and Poirier]{sokalski1983cumulative}
Sokalski,~W.~A.; Poirier,~R. Cumulative atomic multipole representation of the molecular charge distribution and its basis set dependence. \emph{Chemical Physics Letters} \textbf{1983}, \emph{98}, 86--92\relax
\mciteBstWouldAddEndPuncttrue
\mciteSetBstMidEndSepPunct{\mcitedefaultmidpunct}
{\mcitedefaultendpunct}{\mcitedefaultseppunct}\relax
\EndOfBibitem
\bibitem[Aguado and Madden(2003)Aguado, and Madden]{aguado2003ewald}
Aguado,~A.; Madden,~P.~A. Ewald summation of electrostatic multipole interactions up to the quadrupolar level. \emph{The Journal of chemical physics} \textbf{2003}, \emph{119}, 7471--7483\relax
\mciteBstWouldAddEndPuncttrue
\mciteSetBstMidEndSepPunct{\mcitedefaultmidpunct}
{\mcitedefaultendpunct}{\mcitedefaultseppunct}\relax
\EndOfBibitem
\bibitem[Antes and Thiel(1999)Antes, and Thiel]{antes1999adjusted}
Antes,~I.; Thiel,~W. Adjusted connection atoms for combined quantum mechanical and molecular mechanical methods. \emph{The Journal of Physical Chemistry A} \textbf{1999}, \emph{103}, 9290--9295\relax
\mciteBstWouldAddEndPuncttrue
\mciteSetBstMidEndSepPunct{\mcitedefaultmidpunct}
{\mcitedefaultendpunct}{\mcitedefaultseppunct}\relax
\EndOfBibitem
\bibitem[Zhang \latin{et~al.}(1999)Zhang, Lee, and Yang]{zhang1999pseudobond}
Zhang,~Y.; Lee,~T.-S.; Yang,~W. A pseudobond approach to combining quantum mechanical and molecular mechanical methods. \emph{The Journal of chemical physics} \textbf{1999}, \emph{110}, 46--54\relax
\mciteBstWouldAddEndPuncttrue
\mciteSetBstMidEndSepPunct{\mcitedefaultmidpunct}
{\mcitedefaultendpunct}{\mcitedefaultseppunct}\relax
\EndOfBibitem
\bibitem[DiLabio \latin{et~al.}(2002)DiLabio, Hurley, and Christiansen]{dilabio2002simple}
DiLabio,~G.~A.; Hurley,~M.~M.; Christiansen,~P.~A. Simple one-electron quantum capping potentials for use in hybrid QM/MM studies of biological molecules. \emph{The Journal of chemical physics} \textbf{2002}, \emph{116}, 9578--9584\relax
\mciteBstWouldAddEndPuncttrue
\mciteSetBstMidEndSepPunct{\mcitedefaultmidpunct}
{\mcitedefaultendpunct}{\mcitedefaultseppunct}\relax
\EndOfBibitem
\bibitem[Zhang(2005)]{zhang2005improved}
Zhang,~Y. Improved pseudobonds for combined ab initio quantum mechanical/molecular mechanical methods. \emph{The Journal of chemical physics} \textbf{2005}, \emph{122}, 024114\relax
\mciteBstWouldAddEndPuncttrue
\mciteSetBstMidEndSepPunct{\mcitedefaultmidpunct}
{\mcitedefaultendpunct}{\mcitedefaultseppunct}\relax
\EndOfBibitem
\bibitem[Von~Lilienfeld \latin{et~al.}(2005)Von~Lilienfeld, Tavernelli, Rothlisberger, and Sebastiani]{von2005variational}
Von~Lilienfeld,~O.~A.; Tavernelli,~I.; Rothlisberger,~U.; Sebastiani,~D. Variational optimization of effective atom centered potentials for molecular properties. \emph{The Journal of chemical physics} \textbf{2005}, \emph{122}, 014113\relax
\mciteBstWouldAddEndPuncttrue
\mciteSetBstMidEndSepPunct{\mcitedefaultmidpunct}
{\mcitedefaultendpunct}{\mcitedefaultseppunct}\relax
\EndOfBibitem
\bibitem[Slav{\'\i}{\v{c}}ek and Mart{\'\i}nez(2006)Slav{\'\i}{\v{c}}ek, and Mart{\'\i}nez]{slavivcek2006multicentered}
Slav{\'\i}{\v{c}}ek,~P.; Mart{\'\i}nez,~T.~J. Multicentered valence electron effective potentials: A solution to the link atom problem for ground and excited electronic states. \emph{The Journal of chemical physics} \textbf{2006}, \emph{124}, 084107\relax
\mciteBstWouldAddEndPuncttrue
\mciteSetBstMidEndSepPunct{\mcitedefaultmidpunct}
{\mcitedefaultendpunct}{\mcitedefaultseppunct}\relax
\EndOfBibitem
\bibitem[Shao and Kong(2007)Shao, and Kong]{shao2007yinyang}
Shao,~Y.; Kong,~J. YinYang atom: A simple combined ab initio quantum mechanical molecular mechanical model. \emph{The Journal of Physical Chemistry A} \textbf{2007}, \emph{111}, 3661--3671\relax
\mciteBstWouldAddEndPuncttrue
\mciteSetBstMidEndSepPunct{\mcitedefaultmidpunct}
{\mcitedefaultendpunct}{\mcitedefaultseppunct}\relax
\EndOfBibitem
\bibitem[Xiao and Zhang(2007)Xiao, and Zhang]{xiao2007design}
Xiao,~C.; Zhang,~Y. Design-atom approach for the quantum mechanical/molecular mechanical covalent boundary: A design-carbon atom with five valence electrons. \emph{The Journal of chemical physics} \textbf{2007}, \emph{127}, 124102\relax
\mciteBstWouldAddEndPuncttrue
\mciteSetBstMidEndSepPunct{\mcitedefaultmidpunct}
{\mcitedefaultendpunct}{\mcitedefaultseppunct}\relax
\EndOfBibitem
\bibitem[Parks \latin{et~al.}(2008)Parks, Hu, Cohen, and Yang]{parks2008pseudobond}
Parks,~J.~M.; Hu,~H.; Cohen,~A.~J.; Yang,~W. A pseudobond parametrization for improved electrostatics in quantum mechanical/molecular mechanical simulations of enzymes. \emph{The Journal of chemical physics} \textbf{2008}, \emph{129}, 154106\relax
\mciteBstWouldAddEndPuncttrue
\mciteSetBstMidEndSepPunct{\mcitedefaultmidpunct}
{\mcitedefaultendpunct}{\mcitedefaultseppunct}\relax
\EndOfBibitem
\bibitem[Th{\'e}ry \latin{et~al.}(1994)Th{\'e}ry, Rinaldi, Rivail, Maigret, and Ferenczy]{thery1994quantum}
Th{\'e}ry,~V.; Rinaldi,~D.; Rivail,~J.-L.; Maigret,~B.; Ferenczy,~G.~G. Quantum mechanical computations on very large molecular systems: The local self-consistent field method. \emph{Journal of computational chemistry} \textbf{1994}, \emph{15}, 269--282\relax
\mciteBstWouldAddEndPuncttrue
\mciteSetBstMidEndSepPunct{\mcitedefaultmidpunct}
{\mcitedefaultendpunct}{\mcitedefaultseppunct}\relax
\EndOfBibitem
\bibitem[Gao \latin{et~al.}(1998)Gao, Amara, Alhambra, and Field]{gao1998generalized}
Gao,~J.; Amara,~P.; Alhambra,~C.; Field,~M.~J. A generalized hybrid orbital (GHO) method for the treatment of boundary atoms in combined QM/MM calculations. \emph{The Journal of Physical Chemistry A} \textbf{1998}, \emph{102}, 4714--4721\relax
\mciteBstWouldAddEndPuncttrue
\mciteSetBstMidEndSepPunct{\mcitedefaultmidpunct}
{\mcitedefaultendpunct}{\mcitedefaultseppunct}\relax
\EndOfBibitem
\bibitem[Philipp and Friesner(1999)Philipp, and Friesner]{philipp1999mixed}
Philipp,~D.~M.; Friesner,~R.~A. Mixed ab initio QM/MM modeling using frozen orbitals and tests with alanine dipeptide and tetrapeptide. \emph{Journal of computational chemistry} \textbf{1999}, \emph{20}, 1468--1494\relax
\mciteBstWouldAddEndPuncttrue
\mciteSetBstMidEndSepPunct{\mcitedefaultmidpunct}
{\mcitedefaultendpunct}{\mcitedefaultseppunct}\relax
\EndOfBibitem
\bibitem[Sun and Chan(2014)Sun, and Chan]{sun2014exact}
Sun,~Q.; Chan,~G. K.-L. Exact and optimal quantum mechanics/molecular mechanics boundaries. \emph{Journal of chemical theory and computation} \textbf{2014}, \emph{10}, 3784--3790\relax
\mciteBstWouldAddEndPuncttrue
\mciteSetBstMidEndSepPunct{\mcitedefaultmidpunct}
{\mcitedefaultendpunct}{\mcitedefaultseppunct}\relax
\EndOfBibitem
\bibitem[Mardirossian and Head-Gordon(2014)Mardirossian, and Head-Gordon]{mardirossian2014omegab97x}
Mardirossian,~N.; Head-Gordon,~M. $\omega$B97X-V: A 10-parameter, range-separated hybrid, generalized gradient approximation density functional with nonlocal correlation, designed by a survival-of-the-fittest strategy. \emph{Physical Chemistry Chemical Physics} \textbf{2014}, \emph{16}, 9904--9924\relax
\mciteBstWouldAddEndPuncttrue
\mciteSetBstMidEndSepPunct{\mcitedefaultmidpunct}
{\mcitedefaultendpunct}{\mcitedefaultseppunct}\relax
\EndOfBibitem
\bibitem[M{\"u}ller \latin{et~al.}(2023)M{\"u}ller, Hansen, and Grimme]{muller2023omegab97x}
M{\"u}ller,~M.; Hansen,~A.; Grimme,~S. $\omega$B97X-3c: A composite range-separated hybrid DFT method with a molecule-optimized polarized valence double-$\zeta$ basis set. \emph{The Journal of Chemical Physics} \textbf{2023}, \emph{158}, 014103\relax
\mciteBstWouldAddEndPuncttrue
\mciteSetBstMidEndSepPunct{\mcitedefaultmidpunct}
{\mcitedefaultendpunct}{\mcitedefaultseppunct}\relax
\EndOfBibitem
\bibitem[Okuta \latin{et~al.}(2017)Okuta, Unno, Nishino, Hido, and Loomis]{cupy_learningsys2017}
Okuta,~R.; Unno,~Y.; Nishino,~D.; Hido,~S.; Loomis,~C. CuPy: A NumPy-Compatible Library for NVIDIA GPU Calculations. Proceedings of Workshop on Machine Learning Systems (LearningSys) in The Thirty-first Annual Conference on Neural Information Processing Systems (NIPS). 2017\relax
\mciteBstWouldAddEndPuncttrue
\mciteSetBstMidEndSepPunct{\mcitedefaultmidpunct}
{\mcitedefaultendpunct}{\mcitedefaultseppunct}\relax
\EndOfBibitem
\bibitem[Thompson \latin{et~al.}(2022)Thompson, Aktulga, Berger, Bolintineanu, Brown, Crozier, in~'t Veld, Kohlmeyer, Moore, Nguyen, Shan, Stevens, Tranchida, Trott, and Plimpton]{LAMMPS}
Thompson,~A.~P.; Aktulga,~H.~M.; Berger,~R.; Bolintineanu,~D.~S.; Brown,~W.~M.; Crozier,~P.~S.; in~'t Veld,~P.~J.; Kohlmeyer,~A.; Moore,~S.~G.; Nguyen,~T.~D.; Shan,~R.; Stevens,~M.~J.; Tranchida,~J.; Trott,~C.; Plimpton,~S.~J. {LAMMPS} - a flexible simulation tool for particle-based materials modeling at the atomic, meso, and continuum scales. \emph{Comp. Phys. Comm.} \textbf{2022}, \emph{271}, 108171\relax
\mciteBstWouldAddEndPuncttrue
\mciteSetBstMidEndSepPunct{\mcitedefaultmidpunct}
{\mcitedefaultendpunct}{\mcitedefaultseppunct}\relax
\EndOfBibitem
\bibitem[Kapil \latin{et~al.}(2019)Kapil, Rossi, Marsalek, Petraglia, Litman, Spura, Cheng, Cuzzocrea, Mei{\ss}ner, Wilkins, \latin{et~al.} others]{kapil2019pi}
Kapil,~V.; Rossi,~M.; Marsalek,~O.; Petraglia,~R.; Litman,~Y.; Spura,~T.; Cheng,~B.; Cuzzocrea,~A.; Mei{\ss}ner,~R.~H.; Wilkins,~D.~M.; others i-PI 2.0: A universal force engine for advanced molecular simulations. \emph{Computer Physics Communications} \textbf{2019}, \emph{236}, 214--223\relax
\mciteBstWouldAddEndPuncttrue
\mciteSetBstMidEndSepPunct{\mcitedefaultmidpunct}
{\mcitedefaultendpunct}{\mcitedefaultseppunct}\relax
\EndOfBibitem
\bibitem[Pyykk{\"o} and Atsumi(2009)Pyykk{\"o}, and Atsumi]{pyykko2009molecular}
Pyykk{\"o},~P.; Atsumi,~M. Molecular single-bond covalent radii for elements 1--118. \emph{Chemistry--A European Journal} \textbf{2009}, \emph{15}, 186--197\relax
\mciteBstWouldAddEndPuncttrue
\mciteSetBstMidEndSepPunct{\mcitedefaultmidpunct}
{\mcitedefaultendpunct}{\mcitedefaultseppunct}\relax
\EndOfBibitem
\bibitem[Lindbo and Tornberg(2011)Lindbo, and Tornberg]{lindbo2011spectral}
Lindbo,~D.; Tornberg,~A.-K. Spectral accuracy in fast Ewald-based methods for particle simulations. \emph{Journal of Computational Physics} \textbf{2011}, \emph{230}, 8744--8761\relax
\mciteBstWouldAddEndPuncttrue
\mciteSetBstMidEndSepPunct{\mcitedefaultmidpunct}
{\mcitedefaultendpunct}{\mcitedefaultseppunct}\relax
\EndOfBibitem
\bibitem[Tanaka \latin{et~al.}(2013)Tanaka, Katouda, and Nagase]{tanaka2013optimization}
Tanaka,~M.; Katouda,~M.; Nagase,~S. Optimization of RI-MP2 Auxiliary Basis Functions for 6-31G** and 6-311G** Basis Sets for First-, Second-, and Third-Row Elements. \emph{Journal of Computational Chemistry} \textbf{2013}, \emph{34}, 2568--2575\relax
\mciteBstWouldAddEndPuncttrue
\mciteSetBstMidEndSepPunct{\mcitedefaultmidpunct}
{\mcitedefaultendpunct}{\mcitedefaultseppunct}\relax
\EndOfBibitem
\bibitem[Sun \latin{et~al.}(2020)Sun, Zhang, Banerjee, Bao, Barbry, Blunt, Bogdanov, Booth, Chen, Cui, Eriksen, Gao, Guo, Hermann, Hermes, Koh, Koval, Lehtola, Li, Liu, Mardirossian, McClain, Motta, Mussard, Pham, Pulkin, Purwanto, Robinson, Ronca, Sayfutyarova, Scheurer, Schurkus, Smith, Sun, Sun, Upadhyay, Wagner, Wang, White, Whitfield, Williamson, Wouters, Yang, Yu, Zhu, Berkelbach, Sharma, Sokolov, and Chan]{pyscf}
Sun,~Q. \latin{et~al.}  {Recent developments in the PySCF program package}. \emph{The Journal of Chemical Physics} \textbf{2020}, \emph{153}, 024109\relax
\mciteBstWouldAddEndPuncttrue
\mciteSetBstMidEndSepPunct{\mcitedefaultmidpunct}
{\mcitedefaultendpunct}{\mcitedefaultseppunct}\relax
\EndOfBibitem
\bibitem[Kolafa(2004)]{kolafa2004time}
Kolafa,~J. Time-reversible always stable predictor--corrector method for molecular dynamics of polarizable molecules. \emph{Journal of computational chemistry} \textbf{2004}, \emph{25}, 335--342\relax
\mciteBstWouldAddEndPuncttrue
\mciteSetBstMidEndSepPunct{\mcitedefaultmidpunct}
{\mcitedefaultendpunct}{\mcitedefaultseppunct}\relax
\EndOfBibitem
\bibitem[K{\"u}hne \latin{et~al.}(2007)K{\"u}hne, Krack, Mohamed, and Parrinello]{kuhne2007efficient}
K{\"u}hne,~T.~D.; Krack,~M.; Mohamed,~F.~R.; Parrinello,~M. Efficient and accurate Car-Parrinello-like approach to Born-Oppenheimer molecular dynamics. \emph{Physical review letters} \textbf{2007}, \emph{98}, 066401\relax
\mciteBstWouldAddEndPuncttrue
\mciteSetBstMidEndSepPunct{\mcitedefaultmidpunct}
{\mcitedefaultendpunct}{\mcitedefaultseppunct}\relax
\EndOfBibitem
\bibitem[Wu \latin{et~al.}(2006)Wu, Tepper, and Voth]{wu2006flexible}
Wu,~Y.; Tepper,~H.~L.; Voth,~G.~A. Flexible simple point-charge water model with improved liquid-state properties. \emph{The Journal of chemical physics} \textbf{2006}, \emph{124}, 024503\relax
\mciteBstWouldAddEndPuncttrue
\mciteSetBstMidEndSepPunct{\mcitedefaultmidpunct}
{\mcitedefaultendpunct}{\mcitedefaultseppunct}\relax
\EndOfBibitem
\bibitem[Mart{\'\i}nez \latin{et~al.}(2009)Mart{\'\i}nez, Andrade, Birgin, and Mart{\'\i}nez]{martinez2009packmol}
Mart{\'\i}nez,~L.; Andrade,~R.; Birgin,~E.~G.; Mart{\'\i}nez,~J.~M. PACKMOL: A package for building initial configurations for molecular dynamics simulations. \emph{Journal of computational chemistry} \textbf{2009}, \emph{30}, 2157--2164\relax
\mciteBstWouldAddEndPuncttrue
\mciteSetBstMidEndSepPunct{\mcitedefaultmidpunct}
{\mcitedefaultendpunct}{\mcitedefaultseppunct}\relax
\EndOfBibitem
\bibitem[Perdew \latin{et~al.}(1996)Perdew, Burke, and Ernzerhof]{perdew1996generalized}
Perdew,~J.~P.; Burke,~K.; Ernzerhof,~M. Generalized gradient approximation made simple. \emph{Physical review letters} \textbf{1996}, \emph{77}, 3865\relax
\mciteBstWouldAddEndPuncttrue
\mciteSetBstMidEndSepPunct{\mcitedefaultmidpunct}
{\mcitedefaultendpunct}{\mcitedefaultseppunct}\relax
\EndOfBibitem
\bibitem[Beckett and Voth(2023)Beckett, and Voth]{beckett2023unveiling}
Beckett,~D.; Voth,~G.~A. Unveiling the catalytic mechanism of GTP hydrolysis in microtubules. \emph{Proceedings of the National Academy of Sciences} \textbf{2023}, \emph{120}, e2305899120\relax
\mciteBstWouldAddEndPuncttrue
\mciteSetBstMidEndSepPunct{\mcitedefaultmidpunct}
{\mcitedefaultendpunct}{\mcitedefaultseppunct}\relax
\EndOfBibitem
\bibitem[Huang \latin{et~al.}(2017)Huang, Rauscher, Nawrocki, Ran, Feig, De~Groot, Grubm{\"u}ller, and MacKerell~Jr]{huang2017charmm36m}
Huang,~J.; Rauscher,~S.; Nawrocki,~G.; Ran,~T.; Feig,~M.; De~Groot,~B.~L.; Grubm{\"u}ller,~H.; MacKerell~Jr,~A.~D. CHARMM36m: an improved force field for folded and intrinsically disordered proteins. \emph{Nature methods} \textbf{2017}, \emph{14}, 71--73\relax
\mciteBstWouldAddEndPuncttrue
\mciteSetBstMidEndSepPunct{\mcitedefaultmidpunct}
{\mcitedefaultendpunct}{\mcitedefaultseppunct}\relax
\EndOfBibitem
\bibitem[Stephens \latin{et~al.}(1994)Stephens, Devlin, Chabalowski, and Frisch]{stephens1994ab}
Stephens,~P.~J.; Devlin,~F.~J.; Chabalowski,~C.~F.; Frisch,~M.~J. Ab initio calculation of vibrational absorption and circular dichroism spectra using density functional force fields. \emph{The Journal of physical chemistry} \textbf{1994}, \emph{98}, 11623--11627\relax
\mciteBstWouldAddEndPuncttrue
\mciteSetBstMidEndSepPunct{\mcitedefaultmidpunct}
{\mcitedefaultendpunct}{\mcitedefaultseppunct}\relax
\EndOfBibitem
\bibitem[Ray \latin{et~al.}(0)Ray, Das, and Raucci]{ray}
Ray,~D.; Das,~S.; Raucci,~U. Kinetic View of Enzyme Catalysis from Enhanced Sampling QM/MM Simulations. \emph{Journal of Chemical Information and Modeling} \textbf{0}, \emph{0}, null, PMID: 38607669\relax
\mciteBstWouldAddEndPuncttrue
\mciteSetBstMidEndSepPunct{\mcitedefaultmidpunct}
{\mcitedefaultendpunct}{\mcitedefaultseppunct}\relax
\EndOfBibitem
\bibitem[Ray \latin{et~al.}(2022)Ray, Ansari, Rizzi, Invernizzi, and Parrinello]{ray2022rare}
Ray,~D.; Ansari,~N.; Rizzi,~V.; Invernizzi,~M.; Parrinello,~M. Rare event kinetics from adaptive bias enhanced sampling. \emph{Journal of Chemical Theory and Computation} \textbf{2022}, \emph{18}, 6500--6509\relax
\mciteBstWouldAddEndPuncttrue
\mciteSetBstMidEndSepPunct{\mcitedefaultmidpunct}
{\mcitedefaultendpunct}{\mcitedefaultseppunct}\relax
\EndOfBibitem
\bibitem[Tribello \latin{et~al.}(2014)Tribello, Bonomi, Branduardi, Camilloni, and Bussi]{tribello2014plumed}
Tribello,~G.~A.; Bonomi,~M.; Branduardi,~D.; Camilloni,~C.; Bussi,~G. PLUMED 2: New feathers for an old bird. \emph{Computer physics communications} \textbf{2014}, \emph{185}, 604--613\relax
\mciteBstWouldAddEndPuncttrue
\mciteSetBstMidEndSepPunct{\mcitedefaultmidpunct}
{\mcitedefaultendpunct}{\mcitedefaultseppunct}\relax
\EndOfBibitem
\bibitem[Bonomi \latin{et~al.}(2019)Bonomi, Bussi, Camilloni, Tribello, Ban{\'a}{\v{s}}, Barducci, Bernetti, Bolhuis, Bottaro, Branduardi, \latin{et~al.} others]{bonomi2019promoting}
Bonomi,~M.; Bussi,~G.; Camilloni,~C.; Tribello,~G.~A.; Ban{\'a}{\v{s}},~P.; Barducci,~A.; Bernetti,~M.; Bolhuis,~P.~G.; Bottaro,~S.; Branduardi,~D.; others Promoting transparency and reproducibility in enhanced molecular simulations. \emph{Nature methods} \textbf{2019}, \emph{16}, 670--673\relax
\mciteBstWouldAddEndPuncttrue
\mciteSetBstMidEndSepPunct{\mcitedefaultmidpunct}
{\mcitedefaultendpunct}{\mcitedefaultseppunct}\relax
\EndOfBibitem
\bibitem[Smidstrup \latin{et~al.}(2014)Smidstrup, Pedersen, Stokbro, and J{\'o}nsson]{smidstrup2014improved}
Smidstrup,~S.; Pedersen,~A.; Stokbro,~K.; J{\'o}nsson,~H. Improved initial guess for minimum energy path calculations. \emph{The Journal of chemical physics} \textbf{2014}, \emph{140}, 214106\relax
\mciteBstWouldAddEndPuncttrue
\mciteSetBstMidEndSepPunct{\mcitedefaultmidpunct}
{\mcitedefaultendpunct}{\mcitedefaultseppunct}\relax
\EndOfBibitem
\bibitem[Larsen \latin{et~al.}(2017)Larsen, Mortensen, Blomqvist, Castelli, Christensen, Du{\l}ak, Friis, Groves, Hammer, Hargus, \latin{et~al.} others]{larsen2017atomic}
Larsen,~A.~H.; Mortensen,~J.~J.; Blomqvist,~J.; Castelli,~I.~E.; Christensen,~R.; Du{\l}ak,~M.; Friis,~J.; Groves,~M.~N.; Hammer,~B.; Hargus,~C.; others The atomic simulation environment—a Python library for working with atoms. \emph{Journal of Physics: Condensed Matter} \textbf{2017}, \emph{29}, 273002\relax
\mciteBstWouldAddEndPuncttrue
\mciteSetBstMidEndSepPunct{\mcitedefaultmidpunct}
{\mcitedefaultendpunct}{\mcitedefaultseppunct}\relax
\EndOfBibitem
\bibitem[Rolik and K{\'a}llay(2011)Rolik, and K{\'a}llay]{rolik2011general}
Rolik,~Z.; K{\'a}llay,~M. A general-order local coupled-cluster method based on the cluster-in-molecule approach. \emph{The Journal of chemical physics} \textbf{2011}, \emph{135}, 104111\relax
\mciteBstWouldAddEndPuncttrue
\mciteSetBstMidEndSepPunct{\mcitedefaultmidpunct}
{\mcitedefaultendpunct}{\mcitedefaultseppunct}\relax
\EndOfBibitem
\bibitem[Neese and Valeev(2011)Neese, and Valeev]{neese2011revisiting}
Neese,~F.; Valeev,~E.~F. Revisiting the atomic natural orbital approach for basis sets: Robust systematic basis sets for explicitly correlated and conventional correlated ab initio methods? \emph{Journal of chemical theory and computation} \textbf{2011}, \emph{7}, 33--43\relax
\mciteBstWouldAddEndPuncttrue
\mciteSetBstMidEndSepPunct{\mcitedefaultmidpunct}
{\mcitedefaultendpunct}{\mcitedefaultseppunct}\relax
\EndOfBibitem
\bibitem[Zhang \latin{et~al.}(2024)Zhang, Li, Ye, Berkelbach, and Chan]{zhang2024performant}
Zhang,~X.; Li,~C.; Ye,~H.-Z.; Berkelbach,~T.~C.; Chan,~G.~K. Performant automatic differentiation of local coupled cluster theories: Response properties and ab initio molecular dynamics. \emph{The Journal of Chemical Physics} \textbf{2024}, \emph{161}, 014109\relax
\mciteBstWouldAddEndPuncttrue
\mciteSetBstMidEndSepPunct{\mcitedefaultmidpunct}
{\mcitedefaultendpunct}{\mcitedefaultseppunct}\relax
\EndOfBibitem
\bibitem[Zhang and Chan(2022)Zhang, and Chan]{zhang2022differentiable}
Zhang,~X.; Chan,~G.~K. Differentiable quantum chemistry with PySCF for molecules and materials at the mean-field level and beyond. \emph{The Journal of Chemical Physics} \textbf{2022}, \emph{157}, 204801\relax
\mciteBstWouldAddEndPuncttrue
\mciteSetBstMidEndSepPunct{\mcitedefaultmidpunct}
{\mcitedefaultendpunct}{\mcitedefaultseppunct}\relax
\EndOfBibitem
\bibitem[H{\"a}nggi \latin{et~al.}(1990)H{\"a}nggi, Talkner, and Borkovec]{hanggi1990reaction}
H{\"a}nggi,~P.; Talkner,~P.; Borkovec,~M. Reaction-rate theory: fifty years after Kramers. \emph{Reviews of modern physics} \textbf{1990}, \emph{62}, 251\relax
\mciteBstWouldAddEndPuncttrue
\mciteSetBstMidEndSepPunct{\mcitedefaultmidpunct}
{\mcitedefaultendpunct}{\mcitedefaultseppunct}\relax
\EndOfBibitem
\bibitem[Yamaguchi \latin{et~al.}(1994)Yamaguchi, Schaefer, Osamura, Goddard, \latin{et~al.} others]{yamaguchi1994new}
Yamaguchi,~Y.; Schaefer,~H.~F.; Osamura,~Y.; Goddard,~J.; others \emph{A new dimension to quantum chemistry: analytic derivative methods in ab initio molecular electronic structure theory}; Oxford University Press, USA, 1994\relax
\mciteBstWouldAddEndPuncttrue
\mciteSetBstMidEndSepPunct{\mcitedefaultmidpunct}
{\mcitedefaultendpunct}{\mcitedefaultseppunct}\relax
\EndOfBibitem
\bibitem[Cruzeiro \latin{et~al.}(2023)Cruzeiro, Wang, Pieri, Hohenstein, and Mart{\'\i}nez]{cruzeiro2023terachem}
Cruzeiro,~V. W.~D.; Wang,~Y.; Pieri,~E.; Hohenstein,~E.~G.; Mart{\'\i}nez,~T.~J. TeraChem protocol buffers (TCPB): Accelerating QM and QM/MM simulations with a client--server model. \emph{The Journal of Chemical Physics} \textbf{2023}, \emph{158}, 044801\relax
\mciteBstWouldAddEndPuncttrue
\mciteSetBstMidEndSepPunct{\mcitedefaultmidpunct}
{\mcitedefaultendpunct}{\mcitedefaultseppunct}\relax
\EndOfBibitem
\bibitem[Kast \latin{et~al.}(1996)Kast, Asif-Ullah, and Hilvert]{kast1996chorismate}
Kast,~P.; Asif-Ullah,~M.; Hilvert,~D. Is chorismate mutase a prototypic entropy trap?-Activation parameters for the Bacillus subtilis enzyme. \emph{Tetrahedron letters} \textbf{1996}, \emph{37}, 2691--2694\relax
\mciteBstWouldAddEndPuncttrue
\mciteSetBstMidEndSepPunct{\mcitedefaultmidpunct}
{\mcitedefaultendpunct}{\mcitedefaultseppunct}\relax
\EndOfBibitem
\bibitem[Guilford \latin{et~al.}(1987)Guilford, Copley, and Knowles]{guilford1987mechanism}
Guilford,~W.~J.; Copley,~S.~D.; Knowles,~J.~R. The mechanism of the chorismate mutase reaction. \emph{Journal of the American Chemical Society} \textbf{1987}, \emph{109}, 5013--5019\relax
\mciteBstWouldAddEndPuncttrue
\mciteSetBstMidEndSepPunct{\mcitedefaultmidpunct}
{\mcitedefaultendpunct}{\mcitedefaultseppunct}\relax
\EndOfBibitem
\bibitem[Ku \latin{et~al.}(1966)Ku, \latin{et~al.} others]{ku1966notes}
Ku,~H.~H.; others Notes on the use of propagation of error formulas. \emph{Journal of Research of the National Bureau of Standards} \textbf{1966}, \emph{70}\relax
\mciteBstWouldAddEndPuncttrue
\mciteSetBstMidEndSepPunct{\mcitedefaultmidpunct}
{\mcitedefaultendpunct}{\mcitedefaultseppunct}\relax
\EndOfBibitem
\bibitem[Hubrich \latin{et~al.}(2021)Hubrich, M{\"u}ller, and Andexer]{hubrich2021chorismate}
Hubrich,~F.; M{\"u}ller,~M.; Andexer,~J.~N. Chorismate-and isochorismate converting enzymes: Versatile catalysts acting on an important metabolic node. \emph{Chemical Communications} \textbf{2021}, \emph{57}, 2441--2463\relax
\mciteBstWouldAddEndPuncttrue
\mciteSetBstMidEndSepPunct{\mcitedefaultmidpunct}
{\mcitedefaultendpunct}{\mcitedefaultseppunct}\relax
\EndOfBibitem
\bibitem[Tzin and Galili(2010)Tzin, and Galili]{tzin2010new}
Tzin,~V.; Galili,~G. New insights into the shikimate and aromatic amino acids biosynthesis pathways in plants. \emph{Molecular plant} \textbf{2010}, \emph{3}, 956--972\relax
\mciteBstWouldAddEndPuncttrue
\mciteSetBstMidEndSepPunct{\mcitedefaultmidpunct}
{\mcitedefaultendpunct}{\mcitedefaultseppunct}\relax
\EndOfBibitem
\bibitem[Noda and Kondo(2017)Noda, and Kondo]{noda2017recent}
Noda,~S.; Kondo,~A. Recent advances in microbial production of aromatic chemicals and derivatives. \emph{Trends in Biotechnology} \textbf{2017}, \emph{35}, 785--796\relax
\mciteBstWouldAddEndPuncttrue
\mciteSetBstMidEndSepPunct{\mcitedefaultmidpunct}
{\mcitedefaultendpunct}{\mcitedefaultseppunct}\relax
\EndOfBibitem
\bibitem[Grubm{\"u}ller(1995)]{grubmuller1995predicting}
Grubm{\"u}ller,~H. Predicting slow structural transitions in macromolecular systems: Conformational flooding. \emph{Physical Review E} \textbf{1995}, \emph{52}, 2893\relax
\mciteBstWouldAddEndPuncttrue
\mciteSetBstMidEndSepPunct{\mcitedefaultmidpunct}
{\mcitedefaultendpunct}{\mcitedefaultseppunct}\relax
\EndOfBibitem
\bibitem[Chook \latin{et~al.}(1994)Chook, Gray, Ke, and Lipscomb]{chook1994monofunctional}
Chook,~Y.~M.; Gray,~J.~V.; Ke,~H.; Lipscomb,~W.~N. The monofunctional chorismate mutase from Bacillus subtilis: structure determination of chorismate mutase and its complexes with a transition state analog and prephenate, and implications for the mechanism of the enzymatic reaction. \emph{Journal of molecular biology} \textbf{1994}, \emph{240}, 476--500\relax
\mciteBstWouldAddEndPuncttrue
\mciteSetBstMidEndSepPunct{\mcitedefaultmidpunct}
{\mcitedefaultendpunct}{\mcitedefaultseppunct}\relax
\EndOfBibitem
\bibitem[Chook \latin{et~al.}(1993)Chook, Ke, and Lipscomb]{chook1993crystal}
Chook,~Y.~M.; Ke,~H.; Lipscomb,~W.~N. Crystal structures of the monofunctional chorismate mutase from Bacillus subtilis and its complex with a transition state analog. \emph{Proceedings of the National Academy of Sciences} \textbf{1993}, \emph{90}, 8600--8603\relax
\mciteBstWouldAddEndPuncttrue
\mciteSetBstMidEndSepPunct{\mcitedefaultmidpunct}
{\mcitedefaultendpunct}{\mcitedefaultseppunct}\relax
\EndOfBibitem
\bibitem[Kast \latin{et~al.}(1996)Kast, Hartgerink, Asif-Ullah, and Hilvert]{kast1996electrostatic}
Kast,~P.; Hartgerink,~J.~D.; Asif-Ullah,~M.; Hilvert,~D. Electrostatic catalysis of the Claisen rearrangement: Probing the role of Glu78 in Bacillus subtilis chorismate mutase by genetic selection. \emph{Journal of the American Chemical Society} \textbf{1996}, \emph{118}, 3069--3070\relax
\mciteBstWouldAddEndPuncttrue
\mciteSetBstMidEndSepPunct{\mcitedefaultmidpunct}
{\mcitedefaultendpunct}{\mcitedefaultseppunct}\relax
\EndOfBibitem
\bibitem[Liu \latin{et~al.}(1996)Liu, Cload, Pastor, and Schultz]{liu1996analysis}
Liu,~D.~R.; Cload,~S.~T.; Pastor,~R.~M.; Schultz,~P.~G. Analysis of active site residues in Escherichia coli chorismate mutase by site-directed mutagenesis. \emph{Journal of the American Chemical Society} \textbf{1996}, \emph{118}, 1789--1790\relax
\mciteBstWouldAddEndPuncttrue
\mciteSetBstMidEndSepPunct{\mcitedefaultmidpunct}
{\mcitedefaultendpunct}{\mcitedefaultseppunct}\relax
\EndOfBibitem
\bibitem[Burschowsky \latin{et~al.}(2014)Burschowsky, van Eerde, {\"O}kvist, Kienh{\"o}fer, Kast, Hilvert, and Krengel]{burschowsky2014electrostatic}
Burschowsky,~D.; van Eerde,~A.; {\"O}kvist,~M.; Kienh{\"o}fer,~A.; Kast,~P.; Hilvert,~D.; Krengel,~U. Electrostatic transition state stabilization rather than reactant destabilization provides the chemical basis for efficient chorismate mutase catalysis. \emph{Proceedings of the National Academy of Sciences} \textbf{2014}, \emph{111}, 17516--17521\relax
\mciteBstWouldAddEndPuncttrue
\mciteSetBstMidEndSepPunct{\mcitedefaultmidpunct}
{\mcitedefaultendpunct}{\mcitedefaultseppunct}\relax
\EndOfBibitem
\end{mcitethebibliography}


\providecommand{\latin}[1]{#1}
\makeatletter
\providecommand{\doi}
  {\begingroup\let\do\@makeother\dospecials
  \catcode`\{=1 \catcode`\}=2 \doi@aux}
\providecommand{\doi@aux}[1]{\endgroup\texttt{#1}}
\makeatother
\providecommand*\mcitethebibliography{\thebibliography}
\csname @ifundefined\endcsname{endmcitethebibliography}  {\let\endmcitethebibliography\endthebibliography}{}
\begin{mcitethebibliography}{8}
\providecommand*\natexlab[1]{#1}
\providecommand*\mciteSetBstSublistMode[1]{}
\providecommand*\mciteSetBstMaxWidthForm[2]{}
\providecommand*\mciteBstWouldAddEndPuncttrue
  {\def\EndOfBibitem{\unskip.}}
\providecommand*\mciteBstWouldAddEndPunctfalse
  {\let\EndOfBibitem\relax}
\providecommand*\mciteSetBstMidEndSepPunct[3]{}
\providecommand*\mciteSetBstSublistLabelBeginEnd[3]{}
\providecommand*\EndOfBibitem{}
\mciteSetBstSublistMode{f}
\mciteSetBstMaxWidthForm{subitem}{(\alph{mcitesubitemcount})}
\mciteSetBstSublistLabelBeginEnd
  {\mcitemaxwidthsubitemform\space}
  {\relax}
  {\relax}

\bibitem[Aguado and Madden(2003)Aguado, and Madden]{aguado2003ewald}
Aguado,~A.; Madden,~P.~A. Ewald summation of electrostatic multipole interactions up to the quadrupolar level. \emph{The Journal of chemical physics} \textbf{2003}, \emph{119}, 7471--7483\relax
\mciteBstWouldAddEndPuncttrue
\mciteSetBstMidEndSepPunct{\mcitedefaultmidpunct}
{\mcitedefaultendpunct}{\mcitedefaultseppunct}\relax
\EndOfBibitem
\bibitem[Pyykk{\"o} and Atsumi(2009)Pyykk{\"o}, and Atsumi]{pyykko2009molecular}
Pyykk{\"o},~P.; Atsumi,~M. Molecular single-bond covalent radii for elements 1--118. \emph{Chemistry--A European Journal} \textbf{2009}, \emph{15}, 186--197\relax
\mciteBstWouldAddEndPuncttrue
\mciteSetBstMidEndSepPunct{\mcitedefaultmidpunct}
{\mcitedefaultendpunct}{\mcitedefaultseppunct}\relax
\EndOfBibitem
\bibitem[Grubm{\"u}ller(1995)]{grubmuller1995predicting}
Grubm{\"u}ller,~H. Predicting slow structural transitions in macromolecular systems: Conformational flooding. \emph{Physical Review E} \textbf{1995}, \emph{52}, 2893\relax
\mciteBstWouldAddEndPuncttrue
\mciteSetBstMidEndSepPunct{\mcitedefaultmidpunct}
{\mcitedefaultendpunct}{\mcitedefaultseppunct}\relax
\EndOfBibitem
\bibitem[H{\"a}nggi \latin{et~al.}(1990)H{\"a}nggi, Talkner, and Borkovec]{hanggi1990reaction}
H{\"a}nggi,~P.; Talkner,~P.; Borkovec,~M. Reaction-rate theory: fifty years after Kramers. \emph{Reviews of modern physics} \textbf{1990}, \emph{62}, 251\relax
\mciteBstWouldAddEndPuncttrue
\mciteSetBstMidEndSepPunct{\mcitedefaultmidpunct}
{\mcitedefaultendpunct}{\mcitedefaultseppunct}\relax
\EndOfBibitem
\bibitem[Ray \latin{et~al.}(2022)Ray, Ansari, Rizzi, Invernizzi, and Parrinello]{ray2022rare}
Ray,~D.; Ansari,~N.; Rizzi,~V.; Invernizzi,~M.; Parrinello,~M. Rare event kinetics from adaptive bias enhanced sampling. \emph{Journal of Chemical Theory and Computation} \textbf{2022}, \emph{18}, 6500--6509\relax
\mciteBstWouldAddEndPuncttrue
\mciteSetBstMidEndSepPunct{\mcitedefaultmidpunct}
{\mcitedefaultendpunct}{\mcitedefaultseppunct}\relax
\EndOfBibitem
\bibitem[Seabold and Perktold(2010)Seabold, and Perktold]{seabold2010statsmodels}
Seabold,~S.; Perktold,~J. Statsmodels: Econometric and statistical modeling with python. 9th Python in Science Conference. 2010\relax
\mciteBstWouldAddEndPuncttrue
\mciteSetBstMidEndSepPunct{\mcitedefaultmidpunct}
{\mcitedefaultendpunct}{\mcitedefaultseppunct}\relax
\EndOfBibitem
\bibitem[Chodera(2016)]{chodera2016simple}
Chodera,~J.~D. A simple method for automated equilibration detection in molecular simulations. \emph{Journal of chemical theory and computation} \textbf{2016}, \emph{12}, 1799--1805\relax
\mciteBstWouldAddEndPuncttrue
\mciteSetBstMidEndSepPunct{\mcitedefaultmidpunct}
{\mcitedefaultendpunct}{\mcitedefaultseppunct}\relax
\EndOfBibitem
\end{mcitethebibliography}
\end{document}


\section{Mutlipolar Ewald}

We specify the expressions we use for the Mulliken multipoles to resolve some potential ambiguities in defining them:
\begin{align}
    Q_i & = q_i - \sum_{\mu\nu} \delta_{\mu\in i} \gamma_{\nu\mu} \langle\mu|\nu\rangle    \\
    \mu_{i\alpha} & = - \sum_{\mu\nu} \delta_{\mu\in i} \gamma_{\nu\mu} \langle\mu|(r_\alpha - R_{i\alpha})|\nu\rangle    \\
    \theta_{i\alpha\beta} & = - \sum_{\mu\nu} \delta_{\mu\in i} \gamma_{\nu\mu} \langle\mu|(r_\alpha - R_{i\alpha})(r_\beta-R_{i\beta})|\nu\rangle    \\
    \Omega_{i\alpha\beta\gamma} & = - \sum_{\mu\nu} \delta_{\mu\in i} \gamma_{\nu\mu} \langle\mu|(r_\alpha - R_{i\alpha})(r_\beta-R_{i\beta})(r_\gamma-R_{i\gamma})|\nu\rangle    
\end{align}
where $q_i$ is the nuclear charge of atom $i$. As discussed in the main text, we define the point charge multipole interaction tensors to be the Taylor expansion coefficients for the Coulomb potential, i.e.,
\begin{align}
&    T_{ji} = \frac{1}{R_{ji}} \\
&    T_{ji\alpha} = \pdpd{}{R_{j\alpha}} T_{ji} = -\frac{R_{ji\alpha}}{R_{ji}^3} \\
&    T_{ji\alpha\beta} = \frac{1}{2}\pdpd{}{R_{j\beta}} T_{ji\alpha} =  \frac{3R_{ji\alpha}R_{ji\beta}-\delta_{\alpha\beta}R_{ji}^2}{2 R_{ji}^5} \\
&    T_{ji\alpha\beta\gamma} = \frac{1}{3}\pdpd{}{R_{j\gamma}}T_{ji\alpha\beta} = - \frac{5R_{ji\alpha}R_{ji\beta}R_{ji\gamma} - R_{ji}^2(R_{ji\alpha}\delta_{\beta\gamma}+R_{ji\beta}\delta_{\alpha\gamma}+R_{ji\gamma}\delta_{\alpha\beta})}{2 R_{ji}^7}
\end{align}
where $R_{ji}=|\mathbf{R}_j-\mathbf{R}_i|$ and $R_{ji\alpha}=R_{j\alpha}-R_{i\alpha}$. 

Similarly, the interaction tensors between multipoles and Gaussian-screened charges are Taylor expansion coefficients of the potential $\erfc{\kappa_{i}R_{ji}}/R_{ji}=1/R_{ji}-\mathrm{erf}(\kappa_{i}R_{ji})/R_{ji}$. Explicitly,
\begin{align}
&    \hat{T}_{ji} = f_1(R_{ji}) \label{eq:THat0} \\
&    \hat{T}_{ji\alpha} = \pdpd{}{R_{j\alpha}} \hat{T}_{ji} = - R_{ji\alpha}f_3(R_{ji})  \label{eq:THat1}\\
&    \hat{T}_{ji\alpha\beta} = \frac{1}{2} \pdpd{}{R_{j\beta}} \hat{T}_{ji\alpha} = 
    \frac{1}{2}\big(3R_{ji\alpha}R_{ji\beta}-\delta_{\alpha\beta}R_{ji}^2\big) f_5(R_{ji}) + \frac{2}{3}\frac{\kappa_i^3}{\sqrt{\pi}}e^{-\kappa_i^2R_{ji}^2}\delta_{\alpha\beta} \label{eq:THat2}\\
&    \hat{T}_{ji\alpha\beta\gamma} = \frac{1}{3} \pdpd{}{R_{j\gamma}} \hat{T}_{ji\alpha\beta} =
    -\frac{1}{2} \big[5R_{ji\alpha}R_{ji\beta}R_{ji\gamma} - R_{ji}^2(R_{ji\alpha}\delta_{\beta\gamma}+R_{ji\beta}\delta_{\alpha\gamma}+R_{ji\gamma}\delta_{\alpha\beta})\big]f_7(R_{ji}) \nonumber \\
&    ~~~~~~~~~~~~~~~~~~~~~~~~~~~~~ - \frac{4}{15}\frac{\kappa_i^5}{\sqrt{\pi}}e^{-\kappa_i^2R_{ji}^2} (R_{ji\alpha}\delta_{\beta\gamma}+R_{ji\beta}\delta_{\alpha\gamma}+R_{ji\gamma}\delta_{\alpha\beta})  \label{eq:THat3}
\end{align}
with
\begin{align}
&    f_1(R_{ji}) = \frac{\erfc{\kappa_i R_{ji}}}{R_{ji}} \\
&    f_3(R_{ji}) = \frac{1}{R_{ji}^2}\Big(f_1(R_{ji}) + \frac{2\kappa_i}{\sqrt{\pi}}e^{-\kappa_i^2R_{ji}^2} \Big)  \\
&    f_5(R_{ji}) = \frac{1}{R_{ji}^2} \Big(f_3(R_{ji}) + \frac{4}{3} \frac{\kappa_i^3}{\sqrt{\pi}} e^{-\kappa_i^2R_{ji}^2}\Big) \\
&    f_7(R_{ji}) = \frac{1}{R_{ji}^2} \Big(f_5(R_{ji})+\frac{8}{15}\frac{\kappa_i^5}{\sqrt{\pi}}e^{-\kappa_i^2R_{ji}^2}\Big)
\end{align}
where $\mathrm{erf}(\kappa_i R_{ji})/R_{ji}$ is the interaction potential between a point charge and a Gaussian-distributed charge with exponent $\kappa_i$.

To facilitate our discussion of the Ewald method for multipoles, we briefly review the Ewald method for point charges. The Ewald approach aims to compute the infinite lattice sum
\begin{align}
    E=\frac{1}{2}\sum_{\mathbf{L}}\sum_{ij}{}^{'} \frac{q_i q_j}{|\mathbf{R}_j-\mathbf{R}_i+\mathbf{L}|}
\end{align}
where $\mathbf{L}$ indicates a translation vector along the lattice, and the primed sum means that we exclude the $i=j$ terms when $\mathbf{L}=\mathbf{0}$. It is well known such a lattice sum is conditionally convergent, meaning that the sum can converge to any value by varying the summation order. The conditional convergence behavior can be traced to the interactions between each cell's total dipoles. One may thus subtract off and add back the dipole-dipole interactions in the lattice sum to result in an absolutely convergent series plus a dipolar energy, the latter of which becomes surface-dependent. In our implementation, we ignore the surface dipolar term, corresponding to embedding the whole lattice in a perfect conductor, known as the ``tin-foil" boundary condition. The absolutely convergent sum can be further decomposed into a short-range sum, a long-range sum, and a self-interaction correction,
\begin{align}
    E_\mathrm{SR} & = \frac{1}{2} \sum_\mathbf{L}\sum_{ij}{}^{'} q_i q_j \frac{ \erfc{\kappa|\mathbf{R}_j-\mathbf{R}_i+\mathbf{L}|} } { |\mathbf{R}_j-\mathbf{R}_i+\mathbf{L}| } \\
    E_\mathrm{LR} & = \frac{2\pi}{V} \sum_{\mathbf{G}\neq\mathbf{0}} \frac{1}{G^2} e^{-\frac{G^2}{4\kappa^2}} \sum_{ij} q_iq_j e^{-i\mathbf{G}\cdot (\mathbf{R}_j-\mathbf{R}_i)} \label{eq:ewald_lr}\\
    E_\mathrm{self} & = \frac{1}{2} \sum_{ij} q_iq_j \delta_{ij} 
    \lim_{R_{ji}\rightarrow0}
    \Big[
    \frac{1}{R_{ji}} - \frac{\erfc{\kappa R_{ji}}}{R_{ji}}
    \Big] \label{eq:ewald_self}
\end{align}
The decomposition is achieved by subtracting a Gaussian-distributed charge (exponent denoted as $\kappa$) with the same amount of charge and at the same location of the point charge $i$, and adding it back. The interaction between a point charge and a resulting Gaussian-screened charge decays exponentially fast in real space and sums to $E_\mathrm{SR}$. The interactions between point charges and the compensating Gaussian-distributed charges decay exponentially fast in  $k$-space and sum to $E_\mathrm{LR}$. Note that the conditionally convergent dipole-dipole interactions have already been subtracted off when deriving Eq.~\ref{eq:ewald_lr}. The self-interaction correction is to remove the spurious interaction between a point charge with its own compensating Gaussian charge introduced in Eq.~\ref{eq:ewald_lr}. This is achieved by evaluating the electrostatic potential generated by the Gaussian charge at its center, as computed by the limit in Eq.~\ref{eq:ewald_self}. We note that Eq.~\ref{eq:ewald_lr} can be simplified using  $\sum_{ij}q_iq_j\exp{(-i\mathbf{G}\cdot(\mathbf{R}_j-\mathbf{R}_i))}=(\sum_i q_i \cos(\mathbf{G}\cdot\mathbf{R}_i))^2+(\sum_i q_i \sin(\mathbf{G}\cdot\mathbf{R}_i))^2$. Eq.~\ref{eq:ewald_self} can also be written more concisely since the limit evaluates to $-\kappa/\sqrt{\pi}$, but we keep the original forms of $E_\mathrm{LR}$ and $E_\mathrm{self}$ for convenience in the following discussion.

To derive the Ewald method for multipoles, we denote the Ewald sum for point charges as $E_\mathrm{Ewald}=\sum_{ji} q_j q_i \psi_{ji}/2$, and Taylor expand the Ewald potential $\psi_{ji}$. The resulting expansion is the sum of the three components, i.e., $\psi_{ji\cdots}=\psi_{ji\cdots}^\text{SR}+\psi_{ji\cdots}^\text{LR}+\psi_{ji\cdots}^\text{self}$. The short-range Ewald potential expansion is given by setting each $\kappa_i$ in Eqs.~\ref{eq:THat0}-\ref{eq:THat3} to be the Ewald $\kappa$. The long-range potential expansion coefficients are
\begin{align}
&    \psi^\mathrm{LR}_{ji} = \frac{4\pi}{V}\sum_{\mathbf{G}\neq\mathbf{0}}\frac{1}{G^2}e^{-\frac{G^2}{4\kappa^2}} e^{-i\mathbf{G}\cdot(\mathbf{R}_j-\mathbf{R}_i)} \\
&    \psi^\mathrm{LR}_{ji\alpha} = \pdpd{}{R_{j\alpha}}\psi^\mathrm{LR}_{ji}= 
    -i\frac{4\pi}{V}\sum_{\mathbf{G}\neq\mathbf{0}}\frac{1}{G^2}e^{-\frac{G^2}{4\kappa^2}} e^{-i\mathbf{G}\cdot(\mathbf{R}_j-\mathbf{R}_i)} G_{j\alpha} \\
&    \psi^\mathrm{LR}_{ji\alpha\beta} = \frac{1}{2}\pdpd{}{R_{j\beta}}\psi^\mathrm{LR}_{ji\alpha}= 
    -\frac{4\pi}{V}\sum_{\mathbf{G}\neq\mathbf{0}}\frac{1}{G^2}e^{-\frac{G^2}{2\kappa^2}} e^{-i\mathbf{G}\cdot(\mathbf{R}_j-\mathbf{R}_i)} G_{j\alpha} G_{j\beta} \\
&    \psi^\mathrm{LR}_{ji\alpha\beta\gamma} = \frac{1}{3}\pdpd{}{R_{j\gamma}}\psi^\mathrm{LR}_{ji\alpha\beta}= 
    i\frac{2\pi}{3V}\sum_{\mathbf{G}\neq\mathbf{0}}\frac{1}{G^2}e^{-\frac{G^2}{4\kappa^2}} e^{-i\mathbf{G}\cdot(\mathbf{R}_j-\mathbf{R}_i)} G_{j\alpha} G_{j\beta} G_{j\gamma}
\end{align}
The self-interaction expansion coefficients are obtained by first Taylor expanding $T_{ji}-\hat{T}_{ji}$ and then taking the limit of $R_{ij} \to 0$. They can be easily derived by using the expressions for $T_{ji\cdots}$ and $\hat{T}_{ji\cdots}$ and noting that every $1/R_{ji}^{2n+1}-f_{2n+1}(R_{ji})$ has a finite limit for each integer $n\geq0$.\cite{aguado2003ewald} As such, 
\begin{align}
&    \psi^\mathrm{self}_{ji} = -\frac{\kappa}{\sqrt{\pi}}\delta_{ji} \\
&    \psi^\mathrm{self}_{ji\alpha} = 0 \\
&    \psi^\mathrm{self}_{ji\alpha\beta} = -\frac{2}{3}\frac{\kappa^3}{\sqrt{\pi}}\delta_{\alpha\beta}\delta_{ji} \\
&    \psi^\mathrm{self}_{ji\alpha\beta\gamma} = 0
\end{align}

Finally, we note that from the detailed expressions for $T$, $\hat{T}$ and $\psi$ tensors up to the third order, we can obtain the nuclear gradient of these tensors up to the second order, that is the truncation level we adopt in our implementation.

\section{Pseudo-bond Parameterization}
We optimized the pseudo-bond parameters for three DFT/basis combinations: $\omega$B97X-3c, $\omega$B97X-V/6-31G** and $\omega$B97X-V/6-311G**, for all 20 natural amino acids except for glycine and proline, as well as the GTP and GDP molecules. The side-chain of each amino acid (with $\alpha$-carbon saturated by hydrogens), and the tri/di-phosphate group (with 5'-carbon saturated by hydrogens) were energy minimized at each level of quantum theory. For protonable residues (aspartate, glutamate, cysteine, serine, threonine, tyrosine, histidine, and lysine), all the protonated and deprotonated forms were considered. The ground-state energy $E$, nuclear gradient $\mathbf{g}$ (which is zero at the geometry minima), and electron density $\rho(\mathbf{r})$ of each molecule were calculated. The terminal methyl group of each model compound was then replaced by a fluorine atom, whose basis and effective core potential (ECP) were individually optimized for each compound to minimize the following loss:
\begin{align}
    \chi^2 = \sum_i \frac{1}{N_i-1} |\Delta E_i - \frac{1}{n_F}\sum_j \Delta E_j|  + \frac{1}{N_i-1} |\Delta \mathbf{g}_i|  + \frac{\sqrt{ \int_A \mathrm{d}\mathbf{r} |\Delta \rho_i(\mathbf{r})|^2 } }{\int_A \mathrm{d}\mathbf{r} \rho_i(\mathbf{r})}
\end{align}
where $i$ indexes a protonation/phosphorylation form of the compound, $N_i$ is the number of atoms in the fluorine-substituted molecule in its $i$-th form, and $n_F$ is the total number of forms. The $\Delta E$, $\Delta \mathbf{g}$, and $\Delta \rho$ are the energy, gradient, and electron density difference between the fluorine-substituted molecule from its methylated form. For compounds with no protonation/phosphorylation variants, the energy loss is always zero. We restricted the pseudo-bond parameters to be shared by all the variants of the same compound. The electron density integral was evaluated on a uniform grid of 0.1 Bohr resolution with an extra 3 Bohr spacing on both sides of the actual extent of the molecule. The subscript $A$ indicates that the integral was performed in the sphere within the covalent radius\cite{pyykko2009molecular} of each atom except for the methyl group and the fluorine atom and excluding the inner core region defined as the sphere of 0.3 Bohr radius. 

The resulting parameters were tested by geometry optimization of the fluorinated molecules and comparing the pairwise distances to those of the original compounds. For compounds with protonation/phosphorylation variants, the protonation and phosphorylation energies were also compared (this is different from the energy loss due to not normalizating over the number of atoms). The test results are summarized in Table~\ref{tab:ps_errors}, and the resulting parameters are provided in the NWChem format in \path{pseudo_bond_params.zip}.

\begin{table}[H]
    \centering
    \begin{tabular}{ccccccc}
    \hline\hline
    Compound & \multicolumn{3}{c}{Distance maximum deviation} & \multicolumn{3}{c}{Protonation/phosphorylation energy error} \\
    \hline
       Ala &        $3.2\times10^{-5}$& $7.7\times10^{-5}$& $4.0\times10^{-5}$ &                          &                  &                   \\
       Arg &        $5.0\times10^{-3}$& $1.3\times10^{-2}$& $6.6\times10^{-3}$ &                          &                  &                   \\
       Asn &        $5.4\times10^{-3}$& $2.2\times10^{-3}$& $4.7\times10^{-3}$ &                          &                  &                   \\
       Asp &        $2.3\times10^{-2}$& $2.3\times10^{-2}$& $3.1\times10^{-2}$ &        $4.5\times10^{-4}$& $1.5            $&$2.4\times10^{-1}$ \\
      Aspp &        $4.1\times10^{-3}$& $4.2\times10^{-3}$& $8.7\times10^{-3}$ &                          &                  &                   \\
       Cys &        $1.6\times10^{-3}$& $3.8\times10^{-3}$& $2.0\times10^{-2}$ &                          &                  &                   \\
      Cysd &        $3.0\times10^{-3}$& $3.8\times10^{-3}$& $1.3\times10^{-2}$ &        $4.1\times10^{-9}$&$5.2\times10^{-7}$&$2.6\times10^{-8}$ \\
       Gln &        $6.6\times10^{-3}$& $1.5\times10^{-2}$& $7.5\times10^{-3}$ &                          &                  &                   \\
       Glu &        $1.9\times10^{-2}$& $1.8\times10^{-2}$& $2.2\times10^{-2}$ &        $7.6\times10^{-5}$&$1.1\times10^{-1}$&$2.4\times10^{-1}$ \\
      Glup &        $1.2\times10^{-2}$& $1.6\times10^{-2}$& $2.1\times10^{-2}$ &                          &                  &                   \\
       Hsd &        $2.3\times10^{-3}$& $5.4\times10^{-3}$& $6.0\times10^{-3}$ &        $1.3\times10^{-1}$&$3.1\times10^{-3}$&$8.3\times10^{-1}$ \\
       Hse &        $4.1\times10^{-3}$& $6.5\times10^{-3}$& $7.2\times10^{-3}$ &        $1.2\times10^{-1}$&$3.1\times10^{-3}$&$4.1\times10^{-1}$ \\
       Hsp &        $8.2\times10^{-3}$& $6.4\times10^{-3}$& $9.5\times10^{-3}$ &                          &                  &                   \\
       Ile &        $1.1\times10^{-2}$& $9.3\times10^{-3}$& $2.2\times10^{-2}$ &                          &                  &                   \\
       Leu &        $1.1\times10^{-2}$& $1.8\times10^{-2}$& $1.2\times10^{-2}$ &                          &                  &                   \\
       Lys &        $5.7\times10^{-3}$& $6.1\times10^{-3}$& $7.3\times10^{-3}$ &                          &                  &                   \\
      Lysd &        $1.1\times10^{-2}$& $1.1\times10^{-2}$& $1.2\times10^{-2}$ &        $1.4\times10^{-1}$&$4.0\times10^{-1}$&$9.6\times10^{-1}$ \\
       Met &        $8.2\times10^{-3}$& $1.2\times10^{-2}$& $2.0\times10^{-2}$ &                          &                  &                   \\
       Phe &        $4.1\times10^{-3}$& $1.7\times10^{-2}$& $6.2\times10^{-3}$ &                          &                  &                   \\
       Ser &        $1.6\times10^{-2}$& $1.4\times10^{-2}$& $1.6\times10^{-2}$ &                          &                  &                   \\
      Serd &        $1.8\times10^{-2}$& $2.5\times10^{-2}$& $2.8\times10^{-2}$ &        $6.5\times10^{-7}$&$4.5\times10^{-2}$&$2.3\times10^{-2}$ \\
       Thr &        $8.7\times10^{-3}$& $1.4\times10^{-2}$& $1.4\times10^{-2}$ &                          &                  &                   \\
      Thrd &        $2.2\times10^{-2}$& $2.6\times10^{-2}$& $2.8\times10^{-2}$ &        $5.1\times10^{-7}$&$5.5\times10^{-8}$&$7.8\times10^{-2}$ \\
       Trp &        $1.1\times10^{-2}$& $1.2\times10^{-2}$& $1.2\times10^{-2}$ &                          &                  &                   \\
       Tyr &        $9.1\times10^{-3}$& $1.1\times10^{-2}$& $7.2\times10^{-3}$ &                          &                  &                   \\
      Tyrd &        $3.0\times10^{-2}$& $2.3\times10^{-2}$& $2.9\times10^{-2}$ &        $3.3\times10^{-4}$&$2.1\times10^{-1}$&$3.1\times10^{-7}$ \\
       Val &        $9.9\times10^{-3}$& $7.5\times10^{-3}$& $8.4\times10^{-3}$ &                                                                 \\    
       GTP &        $1.0\times10^{-1}$& $2.4\times10^{-2}$& $1.3\times10^{-1}$ &                          &                  &                   \\
       GDP &        $1.2\times10^{-1}$& $2.2\times10^{-2}$& $1.3\times10^{-1}$ &        $8.2\times10^{-1}$&$9.9\times10^{-1}$&$6.6\times10^{-1}$ \\
\hline\hline
    \end{tabular}
    \caption{Test results for the pseudo-bond parameters. The columns are in the order of $\omega$B97X-3c, $\omega$B97X-V/6-31G** and $\omega$B97X-V/6-311G**. Energy errors are given in kcal/mol, and distance errors are given in~\AA. Distance errors are computed only for atom pairs that are within 1.7~\AA~in the methylated molecules. The three letter codes for amino acids are used and names without any suffixes indicate the dominant protonated form in neutral pH. Suffix with d/p represents its deprotonated/protonated form from which the protonated state without d/p can be inferred. For histidine, Hsp indicates its double protonated form, Hsd indicates the single protonated form on the $\delta$-nitrogen and Hse indicates a protonated $\epsilon$-nitrogen. }
    \label{tab:ps_errors}
\end{table}

\section{Chorismate Mutase}

\subsection{Steered Molecular Dynamics for Flipping Arg90}
We performed two stages of SMD to transition the Arg90 residue into a conformation that forms two hydrogen bonds with the chorismate substrate. In the first phase, we applied biases to the distances between the chorismate carboxylic and ester oxygens and the terminal $\eta$-nitrogens of Arg90. We also biased the distance between the chorismate carboxylic oxygen and the $\epsilon$-hydrogen of Arg90. The PLUMED input file for this phase is as follows:
\begin{verbatim}
UNITS LENGTH=A
d1: DISTANCE ATOMS=1425,5669
d2: DISTANCE ATOMS=1428,5670
d3: DISTANCE ATOMS=1423,5669

mr: MOVINGRESTRAINT ...
   ARG=d1,d2,d3
   STEP0=0     AT0=3.0,3.3,4.5  KAPPA0=1000,1000,200
   STEP1=3000  AT1=4.0,2.7,3.1  KAPPA1=1000,1000,200
   STEP2=7500  AT2=5.0,2.7,1.7  KAPPA2=0,200,1000
... mr:
\end{verbatim}
In the second phase, we biased the distances between the chorismate ester oxygen and the Arg90 $\eta$-hydrogen, the chorismate carboxylic oxygen and the Arg90 $\epsilon$-hydrogen, and between the Glu78 carboxylic oxygen and the Arg90 $\eta$-nitrogen. The PLUMED input file for this second phase is as follows:
\begin{verbatim}
UNITS LENGTH=A
d1: DISTANCE ATOMS=1239,1428
d2: DISTANCE ATOMS=1430,5670
d3: DISTANCE ATOMS=1423,5669

mr: MOVINGRESTRAINT ...
   ARG=d1,d2,d3
   STEP0=0     AT0=2.7,2.9,1.7  KAPPA0=1000,1000,1000
   STEP1=2000  AT1=5.0,1.5,1.5  KAPPA1=100,1000,1000
   STEP2=2500  AT2=5.0,1.5,1.5  KAPPA2=0,0,0
... mr:
\end{verbatim}

\subsection{On-the-fly Probability Enhanced Sampling Flooding}
As mentioned in the main text, we determined the OPES flooding parameters based on SMD simulations driving the reactant chorismate into the prephenate product. Specifically, in the PLUMED input for OPES flooding, we set the \texttt{BARRIER} value to be roughly 5 kJ/mol lower than the SMD-estimated free energy barrier and defined the bias excluded region boundary at the point where the SMD-estimated free energy roughly reaches the \texttt{BARRIER} value from the reactant side. The PLUMED input for the 1HB PBE/6-31G** simulation is as follows:
\begin{verbatim}
UNITS LENGTH=A
d1: DISTANCE ATOMS=5663,5675
d2: DISTANCE ATOMS=5670,5671

UPPER_WALLS ARG=d1 AT=+3.60 KAPPA=250.0 EXP=2 LABEL=uwall_C-C5

dd: COMBINE ARG=d1,d2 COEFFICIENTS=1,-1 PERIODIC=NO

chi_exc: CUSTOM ARG=dd FUNC=step(-x+0.8) PERIODIC=NO

opes: OPES_METAD ...
  ARG=dd
  PACE=100
  ADAPTIVE_SIGMA_STRIDE=200
  BARRIER=45
  EXCLUDED_REGION=chi_exc
  TEMP=300.0
  STATE_WFILE=RESTART_INFO
  STATE_WSTRIDE=100
...

COMMITTOR ...
 ARG=dd
 STRIDE=10
 BASIN_LL1=-10.0
 BASIN_UL1=-2.0
... COMMITTOR

PRINT ARG=d1,d2,dd,opes.*,uwall_C-C5.* FILE=COLVAR STRIDE=1

FLUSH STRIDE=1
\end{verbatim}
The detailed parameters for other OPES flooding runs are given in Table~\ref{tab:opes_params}.

\begin{table}[H]
    \centering
    \begin{tabular}{llll}
    \hline
Binding Mode/QM region &  PBE/6-31G**  &  $\omega$B97X-3c & $\omega$B97X-3c (refined)\\
\hline
1HB/S+R90              &    45/0.8     &  110/0.45        & \red{79/0.55}     \\
2HB/S+R90              &    \red{32/0.8}           &   88/0.5  & \red{74/0.58}    \\
2HB/S+R90+R7+E78       &    23/0.8      &   84/0.5         & \red{65/0.56}  \\
    \hline
    \end{tabular} \\
    \caption{OPES flooding parameters. For each table entry, the two numbers are the barrier parameter in kJ/mol and the excluded region boundary in~\AA. }
    \label{tab:opes_params}
\end{table}

\subsection{Rate Calculation and Statistical Analysis}
A flooding simulation\cite{grubmuller1995predicting} adds a biasing potential ($V_f$) to the reactant basin to effectively reduce the reaction free energy barrier and enhance the sampling of rare events. According to classical rate theory\cite{hanggi1990reaction}, the reaction rate constant in an unbiased simulation ($k$) and a flooding simulation ($k_f$) are given by the following expressions:
\begin{align}
    k = \kappa \frac{\langle|\dot{\xi}|\rangle_{\xi^{\neq}}}{2}\frac{\int \mathrm{d}r^{3N} e^{-\beta U}\delta(\xi(r^{3N})-\xi^{\neq})} {\int_R \mathrm{d}r^{3N} e^{-\beta U}}
\end{align}
\begin{align}
    k_f = \kappa_f \frac{\langle|\dot{\xi}|\rangle_{\xi^{\neq},f}}{2}\frac{\int \mathrm{d}r^{3N} e^{-\beta(U+V_f)}\delta(\xi(r^{3N})-\xi^{\neq})} {\int_R \mathrm{d}r^{3N} e^{-\beta(U+V_f)}}
\end{align}
Here, $U$ is the potential energy, $\beta$ is the inverse temperature, $\xi$ is one or more reaction coordinates to define the transition state manifold $\xi^{\neq}$, $\kappa$ is the transmission coefficient, and the $\xi$ velocity ensemble average is performed at the transition state.
Since $V_f$ is only applied in the reactant basin and not around the transition state, we have $\int \mathrm{d}r^{3N} e^{-\beta(U+V_f)}\delta(\xi(r^{3N})-\xi^{\neq}) = \int \mathrm{d}r^{3N} e^{-\beta U}\delta(\xi(r^{3N})-\xi^{\neq})$. If the dynamics around the transition state are further assumed to be unperturbed given the absence of $V_f$ around $\xi^{\neq}$, we have $\kappa_f=\kappa$ and $\langle|\dot{\xi}|\rangle_{\xi^{\neq}}=\langle|\dot{\xi}|\rangle_{\xi^{\neq},f}$. This leads to a simple relation between the biased and unbiased dynamics:
\begin{align}
    \frac{t}{t_f} = \frac{k_f}{k} = \frac{\int_R \mathrm{d}r^{3N} e^{-\beta U}}{\int_R \mathrm{d}r^{3N} e^{-\beta(U+V_f)}}
    = \langle e^{\beta V_f}\rangle_{U+V_f,R}
\end{align}
where $t=1/k$ is the mean first passage time. This provides a straightforward way to extract the unbiased reaction rate constant from the accelerated flooding simulation, without requiring explicit knowledge of the transmission coefficients. 
In the OPES flooding variant, the $V_f$ is periodically updated to fill the reactant basin. The ensemble average is carried out as a time average over the OPES flooding trajectory, shot from the reactant basin and stopped when it reaches the product state, as suggested in Ref.~\cite{ray2022rare}. This assumes that the OPES flooding trajectory samples the equilibrium generated by $U+V_f$ which is not necessarily true due to the finite spacing between two updates of $V_f$. Empirically, OPES flooding underestimates the true unbiased rate constant by 2-3 on most of the reported data in Ref.~\cite{ray2022rare}.

To facilitate the statistical analysis of the sampled reaction rate constants, we assume the rate constant obeys a log-normal distribution, i.e., $\ln{k}$ obeys a normal distribution. This assumption arises from an assumed normal-distributed activation free energy (based on a central limit theorem argument) and the exponential dependency of rate constant on the activation free energy. 
Since each OPES flooding run is initiated from a geometry sampled from a consecutive MD run, the sampled reaction rate constants are generally not independent. We model this auto-correlation by a first-order auto-regressive model [AR(1)], which assumes:
\begin{align}
    \ln k_{i+1} = c + \rho \ln k_{i} + \epsilon_{i+1}
\end{align}
where $c$ and $\rho$ are constant parameters, and $\epsilon_{i+1}$ is a white noise with zero mean and variance $\sigma^2$.
The stationary distribution of AR(1) is a normal distribution $\mathcal{N}(c/(1-\rho), \sigma^2/(1-\rho^2))$, and the time series of $\ln k_i$ is a Gaussian process with a covariance matrix $\Sigma_{ij}=\sigma^2\rho^{|i-j|}/(1-\rho^2) $. In this work, the parameters $c, \rho$ and $\sigma^2$ were obtained via a conditional maximum likelihood estimation as implemented in the statsmodels library\cite{seabold2010statsmodels}. We used a simple unbiased estimator for $\mathbb{E}[k]$ based on $n$ samples:
\begin{align}
    \hat{k} = \frac{1}{n} \sum_{i=1}^n k_i
\end{align}
The variance of this estimator is:
\begin{align}
    \mathrm{Var}(\hat{k}) 
    & = \frac{1}{n^2} \sum_{ij} (k_i - \mathbb{E}[k]) (k_j - \mathbb{E}[k]) \nonumber \\
    & = e^{2 M + S^2} \Big(\frac{1}{n^2}\sum_{ij}e^{S^2 \rho^{|i-j|}} - 1 \Big)
\end{align}
where $M$ and $S^2$ are the mean and variance of $\ln{k}$, i.e.,
\begin{align}
    M = \frac{c}{1-\rho} \\
    S^2 = \frac{\sigma^2}{1-\rho^2}
\end{align}
For each combination of Arg90 conformation, quantum theory, and QM region definition, we performed 11 OPES flooding simulations, starting from configurations sampled every 1 ps after 2 ps of initial equilibration. To balance the bias and variance of the rate constant estimate, we discarded the first few data points from our statistics. We determined the optimal number of discarded points by minimizing the variance of the resulting $n$-sample estimator, similar to the approach of Chodera\cite{chodera2016simple} by maximizing the effective sample size. The number of data points kept for the statistics in each condition is summarized in Table~\ref{tab:nsample}.

\begin{table}[H]
    \centering
    \begin{tabular}{llll}
    \hline
     Binding Mode/QM region &  PBE/6-31G**  &  $\omega$B97X-3c & $\omega$B97X-3c (refined)\\
\hline
1HB/S+R90   &  11       & 9   &  \red{11}   \\
2HB/S+R90   &   \red{8}  &   11  &      \red{8}\\
2HB/S+R90+R7+E78  &  9    &    7   &  \red{11}\\
    \hline
    \end{tabular} \\
    \caption{The number of last data points (simulations) used for computing the mean and the variance of the catalytic rate constant of the chorismate mutase.}
    \label{tab:nsample}
\end{table}

\bibliography{ref}